\documentclass[12pt,preprint]{aastex}

\newcommand\simgt{\lower.5ex\hbox{$\; \buildrel > \over \sim \;$}}
\newcommand\simlt{\lower.5ex\hbox{$\; \buildrel < \over \sim \;$}}

\newcommand\Msun{M_\odot}
\newcommand\kms{{\rm\;km\;s^{-1}}}
\newcommand\nJ{n_{\rm J}}
\newcommand\torb{t_{\rm orb}}
\newcommand\Nptl{N_{\rm ptl}}
\newcommand\Qgcrit{Q_{g, \rm crit}}
\newcommand\Qgth{Q_{g, \rm th}}

\newcommand\kscrit{k_{s, \rm crit}}
\newcommand\lmarg{\lambda_{\rm marg}}
\newcommand\dEk{\delta E_{\rm kin}}

\newcommand\sles{\lower2pt\hbox{$\buildrel {\scriptstyle <}
   \over {\scriptstyle\sim}$}}
\newcommand\sgreat{\lower2pt\hbox{$\buildrel {\scriptstyle >}
   \over {\scriptstyle\sim}$}}
\shortauthors{Kim \& Ostriker}
\shorttitle{Star-Gas Galactic Disks}

\begin{document}

\title{Gravitational Runaway and Turbulence Driving in Star-Gas Galactic Disks}

\author{Woong-Tae Kim$^{1}$ and Eve C. Ostriker$^{2}$}
\affil{$^1$Department of Physics and Astronomy, FPRD, 
Seoul National University, Seoul 151-742, Republic of Korea}
\affil{$^2$Department of Astronomy, University of Maryland \\
College Park, MD 20742} 

\email{wkim@astro.snu.ac.kr, ostriker@astro.umd.edu}

\begin{abstract}
Galactic disks consist of both stars and gas.  The gas is more
dynamically responsive than the stars, and strongly nonlinear
structures and velocities can develop in the interstellar medium (ISM)
even while stellar surface density perturbations remain fractionally
small.  Yet, the stellar component still significantly influences the gas.
We use two-dimensional numerical simulations to explore formation of bound
condensations and turbulence generation in the gas of two-component
galactic disks.  We represent the stars with collisionless particles
and follow their orbits using a particle-mesh method, and treat the
gas as an isothermal, unmagnetized fluid.  The two components interact
through a combined gravitational potential, which accounts for the
distinct vertical thickness of each disk.  We ensure our
results are not contaminated by particle Poisson noise, which
can seriously compromise simulations of self-gravitating
systems. Using stellar parameters typical of mid-disk
conditions, we find that models with gaseous Toomre parameter
$Q_g<\Qgcrit \sim 1.4 $ experience gravitational runaway and
eventually form bound condensations.  This $\Qgcrit$ value is nearly
the same as previously found for razor-thin, gas-only models,
indicating that the destabilizing effect of live stars is offsets
the reduced self-gravity of thick disks.  
This result is also consistent with empirical studies showing that star
formation is suppressed when $Q_g\simgt 1-2$.
The bound gaseous structures that
form have mass $6\times 10^7\Msun$ each, $\sim 10$ times larger than
the thin-disk Jeans mass for a gas-only disk; these
represent superclouds that would subsequently fragment into GMCs.
Self-gravity and sheared rotation also interact to drive turbulence in
the gas when $Q_g \simgt \Qgcrit$.  This turbulence is anisotropic,
with more power in sheared than compressive motions.  The gaseous
velocity dispersion is $\sim0.6$ times the thermal speed when
$Q_g\approx \Qgcrit$.  This suggests that gravity is important in
driving ISM turbulence in many spiral galaxies, since the low
efficiency of star formation naturally leads to a state of marginal
instability.
\end{abstract}


\section{Introduction}

Large-scale galactic disk evolution is governed by the interaction
between self-gravity and other reinforcing and/or countervailing
forces. Some of these forces are associated with sheared
galactocentric rotation, and others are associated with random
velocities (thermal and turbulent) of the gas and stars.  Under
certain circumstances, gravitational instabilities may grow and
radically alter the disk structure as they develop nonlinearly.  This
is believed to be how giant molecular clouds (GMCs), and subsequently
star-forming \ion{H}{2}  regions, originate.  Under other circumstances, the
overall disk morphology is not changed, but self-gravitating modes
grow and interact in such a way that significant energy is transfered
from ordered to disordered forms.  This ``disk heating'' may in turn
affect the system's subsequent susceptibility to self-gravitating
instabilities.

Because the gas component is much cooler than that of the stars, and
because gas can radiate energy as it nonlinearly compressed, the
effects of self-gravitating instabilities are much more pronounced in
the interstellar medium (ISM) than in the stellar disk.  The stellar
component can nevertheless be quite important to the initial growth of
self-gravitating modes (or waves), because it contains so much of the
disk's mass ($\sim 75-90\%$, compared to $\sim 10-25\%$ in gas).
While the stars are essentially collisionless, most previous work on
gravitational instability in two-component disks has treated both
stars and gas as two isothermal fluids with different sound speeds.
This approach was taken by \citet{jog84a,jog84b}, \citet{elm95b}, and
\citet{jog96} in studying axisymmetric self-gravitating modes, and by
\citet{jog92} in studying non-axisymmetric self-gravitating modes, for
combined star/gas disks. Adopting a full kinetic treatment for the
stellar component in two-component disks, on the other hand,
\citet{lin69} and \citet{raf01} used linear perturbation theory to
derive dispersion relations for non-axisymmetric WKB waves and
axisymmetric modes, respectively.  These linear-theory analyses cannot
capture, however, the ultimate outcomes -- including GMC formation and
turbulent driving -- which result when localized, self-gravitating
structures grow; quantifying this development requires
nonlinear numerical simulations.

The main class of self-gravitating instability that develops under
conditions of strong sheared rotation (in outer galaxies, away from
spiral arms) is the swing amplifier \citep{gol65b,jul66,too81}.  In
\citet{kim01} (hereafter Paper I), we identified the basic
requirements for swing to occur, and reported on the results of
extensive numerical simulations of swing amplification in gaseous
disks with or without magnetic fields.  Although in linear theory the
amplification magnitude varies continuously with respect to the
Toomre parameter $Q_g$ (see eq.\ [\ref{TQ}] for definition), Paper I
showed that nonlinear interactions among swing-amplified density
filaments or wavelets lead to a threshold phenomenon.  Disks with $Q_g
> \Qgcrit$ remain stable, while disks with $Q_g < \Qgcrit$ experience
gravitational runaway, forming bound condensations.  For the
razor-thin disk models of Paper I, we found that $\Qgcrit\sim 1.2-1.4$,
with the largest value resulting from models with subthermal magnetic
fields.  Given the similarity between the numerically-obtained $\Qgcrit$
values and observationally inferred thresholds for active
star formation in external galaxies \citep{qui72,ken89,cal91,mar01,won02}, 
Paper I supported the notion that self-gravitating instabilities define the
star-formation ``edges'' of disks.

While successfully demonstrating the nonlinear threshold behavior of
the swing amplifier, and quantitatively yielding good agreement with
observations, the models considered in Paper I suffered from two
important drawbacks. First, by assuming infinitesimally thin disks,
they overestimated self-gravity, which tends to overestimate
$\Qgcrit$; second, they did not allow for a dynamically-active stellar
disk, which tends to underestimate $\Qgcrit$.  Subsequent simulations
of pure-gas disks in three dimensions confirmed the basic nonlinear
threshold behavior, but also showed that for realistic disk
temperatures the nonzero thickness of the disk can indeed significantly
lower the value of $\Qgcrit$ to $<1$ \citep{kos02}.  

In this paper, we address the second limitation of Paper I by studying
nonlinear evolution of gravitational instabilities in {\it two-component}
galactic disks, consisting of both stars and gas.  We distribute
collisionless particles to represent stars and follow their orbits
using a direct $N$-body method, while adopting a hydrodynamical
approach for gas.  The two components are allowed to 
interact with each other through the combined gravitational potential
(calculated on a single mesh).  We also allow for nonzero (and
differing) vertical thickness of both stellar and gaseous disks in
solving the Poisson equation for each component.

Given that finite thickness effects tend to reduce $\Qgcrit$
values below observed star formation thresholds in gas-only models, 
a compelling question
is whether allowing for an active stellar disk compensates for this
effect.  Thus, one of our primary objectives is to quantify the
critical $Q_g$ value for gravitational runaway when both a live
stellar component and a thick disk are explicitly considered.
A second important goal is to understand what happens when $Q_g$ is
sufficiently large that gravitational runaway does {\it not} occur.
In particular, the large-scale flows that are driven when gravity is
strong (but not too strong) may be a key to powering turbulence
within the disk.

Many processes may contribute to turbulence in the diffuse
gas.  In addition to self-gravitating instabilities, other candidate
mechanisms that have been proposed include thermal instabilities, the
Parker and magnetorotational instabilities, dynamical instabilities in spiral
shocks, stellar winds, and expanding stellar \ion{H}{2} regions and
supernova blast waves (see \citealt{mac04,elm04} for recent reviews).
Traditionally, stellar processes have been presumed to be the
dominant turbulence drivers, but the lack of observed spatial
correlation between velocity dispersions of \ion{H}{1} gas and
regions of high mass star formation \citep{dic_etal90,vanz99} argues
that other sources must be important as well. In particular, \citet{elm02} and
\citet{elm03} have suggested that the swing amplifier generates
large-scale motions that cascade to produce turbulence and structure on 
small scales as well as large scales.
Initial numerical studies of this process have been pursued by
\citet{wad02}, who performed two-dimensional, thin-disk galactic
simulations with ISM cooling.  They found that a cloudy medium
develops, with a velocity dispersion $\sigma_v \sim 2-5\kms$.  On
the other hand, \citet{kim03} found via three-dimensional isothermal
simulations, with sound speed $c_g=7\kms$ and $Q_g\simgt1.7$, that
turbulence was driven to levels less than $1\kms$.  Based on just these
few models, it has not yet been clear how important
``swing-driven'' turbulence is more generally.

In the present work, we determine how the turbulent velocity amplitude
of the gas, $\sigma_v$, varies as a function of $Q_g$, and also assess
how the presence of a live stellar component affects $\sigma_v$.  By
running a number of models with a range of particle numbers, and also
performing control models with ``passively-evolving'' stars, we are
able to subtract out the contribution to $\sigma_v$ due to Poisson
noise in the particle distribution.  We show that quite high particle
resolution is in fact necessary for accurate determination of $\sigma_v$ 
in the case when $Q_g$ is near $\Qgcrit$; this is of practical importance 
because real disks naturally evolve toward near-critical states.

This paper is organized as follows: In \S2, we present the basic
equations we solve for gas and stars, introduce ``thick-disk''
gravity, and describe our model parameters.  In \S3, we revisit the
linear theory for axisymmetric gravitational instability in
two-component disks with an emphasis on the stabilizing effect of
nonzero disk thickness (omitted in previous studies).  In \S4, we
describe numerical methods we employ for our simulations, and present
results of code tests.  We demonstrate that large Poisson noise in the
stellar particle distribution can lead to spurious results for
$N$-body systems that are nearly or marginally gravitationally unstable.
In \S5, we report on our simulations of two-component disks.  We
present numerical results on $Q_g$ thresholds for (non-axisymmetric)
gravitational runaway, and quantify the overall level and spectral
properties of gravity-driven turbulence when $Q_g\simgt \Qgcrit$
Finally, we summarize our results and discuss their astronomical
implications in \S6.

\section{Equations and Model Parameters}

\subsection{Basic Equations and Thick-Disk Gravity}\label{selfg}

In this paper we investigate the dynamics of local galactic 
disks composed of gas and stars.  We follow the 
evolution of the gas by solving the hydrodynamic equations, while employing 
an $N$-body method for the stars.  
The gas is assumed to be isothermal and unmagnetized, and we
assume that all physical variables other than the gravitational potential 
are independent of the vertical height. 
As in Paper I, we consider a patch of the disk orbiting the 
galaxy with a fixed angular frequency 
$\mathbf{\Omega_0}=\Omega_0\hat{\mathbf z}$, 
and set up a local Cartesian frame with $x$ and $y$ referring to 
radial and azimuthal coordinates, respectively.   We consider
a simulation box with size $L\times L$. 
The equilibrium background velocity arising from galactic rotation 
relative to the center of the box at $x=y=0$ is given 
by ${\mathbf v}_0 = -q\Omega_0 x \hat{\mathbf y}$, 
where $q\equiv -d\ln\Omega/d\ln R$ denotes the dimensionless local shear rate. 
The basic dynamical equations we solve for the gaseous part in the
local frame are identical to those presented in Paper I, except that 
we adopt here a ``thick-disk'' gravitational kernel
in solving the Poisson equation (see below).

To follow the evolution of the stellar component, we use collisionless 
particles.  Restricting particle motions to be in-plane, the shearing-sheet 
equations of motion relative to the center of the box are given by
\begin{equation}\label{eom}
\ddot{\mathbf r} = 
         2q\Omega_0^2 x \mathbf{\hat{x}} - 2\mathbf{\Omega_0}\times
        \dot \mathbf{r}
       - \nabla(\Phi_g + \Phi_s) ,
\end{equation}
where $\mathbf r= x \mathbf{\hat{x}} + y \mathbf{\hat{y}}$ is the 
position vector of a particle,  the dots denote derivatives with
respect to time (e.g., \citealt{jul66,wis88}), and
$\Phi_g$ and $\Phi_s$ are the gravitational potentials from gas and stars,
respectively.
The spatial distribution of the particles at any given time will give
the stellar surface density $\Sigma_s$, which in turn yields $\Phi_s$ via
\begin{equation}\label{Pstar}
   \nabla^2\Phi_s = 4\pi G \Sigma_s h_s(z).
\end{equation}
Here, $h_s(z)$ represents the vertical distribution of the stellar
density and satisfies the normalization condition
$\int_{-\infty}^{\infty}h_s(z)dz=1$.  The Poisson equation for the gas
takes the same form as equation (\ref{Pstar}) with subscripts
changed appropriately.

Studies of disk dynamics often assume $h(z)=\delta(z)$ for simplicity.
This thin-disk approximation would be valid as long as perturbations of
interest do not critically rely on the vertical dimension and
their wavelengths $\lambda$ are much larger than the disk scale height $H$.
For waves with $\lambda$ approaching $2\pi H$, however, the thin-disk
approximation overestimates self-gravity at the disk midplane.
This can be particularly severe in a two-component system in which the 
characteristic length scales and disk scale heights for gas and stars 
are quite different from each other.  
For instance, the critical wavelengths for axisymmetric 
gravitational instability under solar neighborhood conditions are
$\sim10$ kpc for stars (e.g., \citealt{bin87}) and $\sim 1$ kpc for gas 
(e.g., \citealt{elm95a}; Paper I).  
The stellar scale height $H_s \sim 330$ pc 
\citep{che01} is non-negligible compared with
the wavelength of the most vulnerable modes in the combined system. 
Moreover, as perturbations grow via swing amplification they form thin 
overdense filaments with widths comparable to the gas disk's 
scale height (Paper I; see also \S\ref{nswing}), so that the thin-disk
gravity may potentially lead to spurious fragmentation.
It is thus desirable to allow for finite disk thickness in solving
equation (\ref{Pstar}).

Suppose that $h_s(z)$ does not vary with time and that 
$\Sigma_{s}$ is periodic in $x$ and $y$ (or linear combinations of
coordinates; see \S 4.1).
Using the Green's function 
method, one can show that equation (\ref{Pstar})
yields the solution at $z=0$ (omitting the subscript $s$)
\begin{equation}\label{Fou1}
\Phi_{}(\mathbf k) = - \frac{2\pi G \Sigma_{}(\mathbf k)}{|{\mathbf k}|}
\int_{-\infty}^{\infty} h_{}(z') e^{-|{\mathbf k}z'|}dz',
\end{equation}
where $\Phi_{}(\mathbf k)$ and $\Sigma_{ }(\mathbf k)$ refer to the 
Fourier components with wavevector $\mathbf k$. 
When $h_{ }(z)=e^{-|z|/H_{ }}/(2H_{ })$, equation (\ref{Fou1}) simply 
becomes 
\begin{equation}\label{thick}
\Phi_{ }(\mathbf k) = -\frac{2\pi G \Sigma_{ }(\mathbf k)}
{|{\mathbf k}|(1 + |{\mathbf k}| H_{ })},
\end{equation}
which we refer to as ``thick-disk'' gravity, as opposed to thin-disk
gravity for which $H_{ }$ is set equal to zero.  This result was
obtained through a different technique by \citet{elm87} and has been
shown to be quite a good approximation for the reduction of
self-gravity in disks with various density profiles (e.g.,
\citealt{rom92,kos02}).  By direct numeral integration, one can see
that equation (\ref{thick}) slightly underestimates the real potential
for purely self-gravitating disks, but not more than
$14\%$.\footnote{When equation (\ref{thick}) is used to obtain an
  axisymmetric dispersion relation for a disk with $H_g=c_g^2/(\pi G
  \Sigma_g)$ (i.e. $H_g$ determined by gas self-gravity alone), the
  resulting $Q_{\rm crit}= 0.65$ is very close to the value $Q_{\rm crit}=
  0.68$ obtained by \citet{gol65a} as an exact criterion for
  axisymmetric instability.}  \citet{kim06} further found that when
equation (\ref{thick}) is used to calculate self-gravity, simulations
of gas flow across two-dimensional spiral shocks produce results very
similar to fully three-dimensional models.  In what follows, we shall
use equation (\ref{thick}) to compute the self-gravity of gas and stars
with scale heights $H_g$ and $H_s$, respectively.

\subsection{Model Parameters}

As initial models we consider a composite disk of gas and stars with
uniform surface densities $\Sigma_{g0}$ and $\Sigma_{s0}$.
The gaseous disk is set to remain isothermal throughout the entire
evolution with an effective speed of sound $c_g=7\kms$, 
corresponding to
a mean Galactic thermal pressure $P/k\sim 2000-4000$ K cm$^{-3}$
\citep{hei01} and mean midplane density $n_{\rm H}\sim 0.6$ cm$^{-3}$
\citep{dic90}, similar to solar neighborhood values.  
For the stellar disk, we initially distribute 
particles according to the Schwarzschild distribution function
\begin{equation}\label{dist}
f(v_x,v_y) 
= \frac{\Sigma_{s0}}{2\pi \sigma_{s,x} \sigma_{s,y}}
\exp{\left[-\frac{v_x^2}{2\sigma_{s,x}^2} 
           -\frac{(v_y-v_{0,y})^2}{2\sigma_{s,y}^2}\right]},
\end{equation}
where $\sigma_{s,x}$ and $\sigma_{s,y}$ are the
$x$- and $y$-components of the particle velocity dispersion, respectively.  
The values of $\sigma_{s,x}$ and $\sigma_{s,y}$ will vary over time 
as the particles respond to the perturbed gravitational potential.  
Note that
$\Sigma_{s0}=\int f dv_x dv_y$ and $\sigma_{s,y}/\sigma_{s,x} = (1-q/2)^{1/2}$
in equilibrium.  We initially adopt
$\sigma_{s,x}=30\kms$ and $\sigma_{s,y}=2^{-1/2}\sigma_{s,x}$ corresponding to
solar neighborhood conditions with a flat ($q=1$)
rotation curve \citep{bin87}.

When written in dimensionless form, the basic dynamical  equations 
are characterized by the following parameters
\begin{equation}\label{TQ}
Q_g = \frac{\kappa_0 c_g}{\pi G \Sigma_{g0}}, \;\;\;\;\;\;
Q_s = \frac{\kappa_0 \sigma_{s,x}}{3.36 G \Sigma_{s0}},
\end{equation}
\begin{equation}\label{nX}
\nJ = \frac{G \Sigma_{g0} L}{c_g^2},\;\;\;\;\;\;
X=\frac{\kappa_0^2 L}{4\pi^2 G \Sigma_{s0}},
\end{equation}
together with $H_g/L$ and $H_s/L$. 
The Toomre $Q$ parameters defined by equation 
(\ref{TQ}) give the surface densities relative to the critical
values at $Q_g=Q_s=1$ for axisymmetric gravitational instability
in a razor-thin, gas-only or star-only disk \citep{too64,bin87}.
In equations (\ref{nX}), the $\nJ$ ($X$) parameter is the ratio of the 
simulation box size $L$ to the critical wavelength 
$\lambda_{g, \rm crit}=c_g^2/G\Sigma_{g0}$ 
($\lambda_{s, \rm crit}=4\pi^2 G \Sigma_{s0}/\kappa_0^2$)
for axisymmetric gravitational instability in a razor-thin,
non-rotating gaseous (rotating, cold stellar) disk.

Since the parameter space is very large and since we are primarily
interested in the dynamical evolution of the gaseous component in
combined disks, we vary only $Q_g$ (or equivalently $\Sigma_{g0}$)
and fix all the other disk properties.  
For the stellar part, we adopt solar neighborhood values
$\kappa_0=36$ km s$^{-1}$ kpc$^{-1}$ \citep{bin87}, 
$\Sigma_{s0}=35\;\Msun$ pc$^{-2}$ 
\citep{kui89}, and $H_s=$ 330 pc \citep{che01,kar04}. This 
gives $Q_s=2.1$,  
so that in our model disks the stellar component {\it alone},
even in the razor-thin limit, would be immune
to axisymmetric instability. 
We adopt $H_g=170$ pc for the effective scale height of the 
local ISM (e.g., \citealt{bou90}).  
For the box size, we adopt $X=2$ or $L=9.2$ kpc,
which is large enough to resolve the most susceptible modes
of the swing amplifier in stellar disks \citep{too81};
test simulations we have performed confirm that results
are insensitive to $X$ as long as $X\geq2$.

Our fiducial model has $Q_g=1.4$ corresponding to 
$\Sigma_{g0}=13\; \Msun$ pc$^{-2}$ for the gas disk
(e.g., \citealt{hol00}), 
but we also allow for various values of $Q_g$ in the range
of 1 -- 3.  These models will allow us to determine the critical $Q_g$
value for gravitational runaway and to study the characteristics 
and level of turbulence driven by self-gravity when $Q_g\simgt \Qgcrit$.
Since equations (\ref{TQ}) and (\ref{nX}) give 
$\nJ=3.7 X\sigma_{s,x}/(c_gQ_g Q_s) \sim 15/Q_g$ for the adopted
set of parameters, we see that our simulation box is large
enough to contain at least 5 Jeans wavelengths of the gaseous medium.
We define the orbital period $\torb \equiv 2\pi/\Omega_0 = 2.4\times 10^8\;
{\rm yr}\; (\Omega_0/26\;{\rm km\;s^{-1}\;kpc^{-1}})^{-1}$
and use it as the time unit in our presentation.

\section{Axisymmetric Stability}\label{linear}

The gravitational stability of two-component disks to axisymmetric
perturbations was first analyzed by \citet{jog84a}, \citet{rom92},
\citet{elm95b}, and \citet{jog96}, who treated the stellar disk as an
isothermal fluid.  While a fluid description of stellar particles is
a good approximation for a range of wavenumbers in disks near the
threshold, it generally fails when disks are further from instability
or when the wavelengths of perturbations are small compared to the
epicyclic excursions of stars (e.g., \citealt{bin87}).  \citet{raf01}
instead used a collisionless description of the stellar population and
derived a dispersion relation for axisymmetric waves in star-gas disks
assuming that the disk is razor-thin.  In this section, we revisit
axisymmetric instability with a particular emphasis on the effect of
nonzero disk thickness.  The results of this section will be used to
check our numerical technique in \S\ref{test}.

We refer the reader to \citet{raf01} for a detailed derivation leading
to the thin-disk dispersion relation for two-component disks.  We
follow the same steps, except we use thick-disk equation (\ref{thick})
for the potential and density pairs in place of Rafikov's equation
(14).  The resulting dispersion relation for axisymmetric modes in a
two-component disk of finite vertical extent reads
\begin{equation}\label{disp}
\frac{2\pi G k \Sigma_{g0}}
     {(\kappa^2 + k^2c_g^2 - \omega^2)(1 + k H_g)}          +
\frac{2\pi G k \Sigma_{s0}
          {\mathcal F}(\omega/\kappa,k^2\sigma_{s,x}^2/\kappa^2)}
     {(\kappa^2 - \omega^2)(1 + k H_s)}=1,
\end{equation}
where $\omega$ and $k$ are the frequency and radial wavenumber
of perturbations, respectively, and ${\mathcal F}$ is the stellar ``reduction 
factor'' defined by equation (6-45) of \citet{bin87}, which accounts for 
the effects of the stellar velocity dispersion.  A similar relation
was presented in \citet{rom92} who introduced a single effective scale
height for the combined disk instead of using different heights
$H_g$ and $H_s$ for each component.  In the limit
of $H_g=H_s=0$, equation (\ref{disp}) recovers the thin-disk
dispersion relation of \citet{raf01}.  Note that although
a combined thick disk can be mapped to a razor-thin 
counterpart with lower surface density,  the density reduction factors
of the gaseous and stellar components differ from each other and 
cannot be determined
a priori since they depend on the perturbation wavenumber.  
 
\begin{figure}
\epsscale{1.00}
\plotone{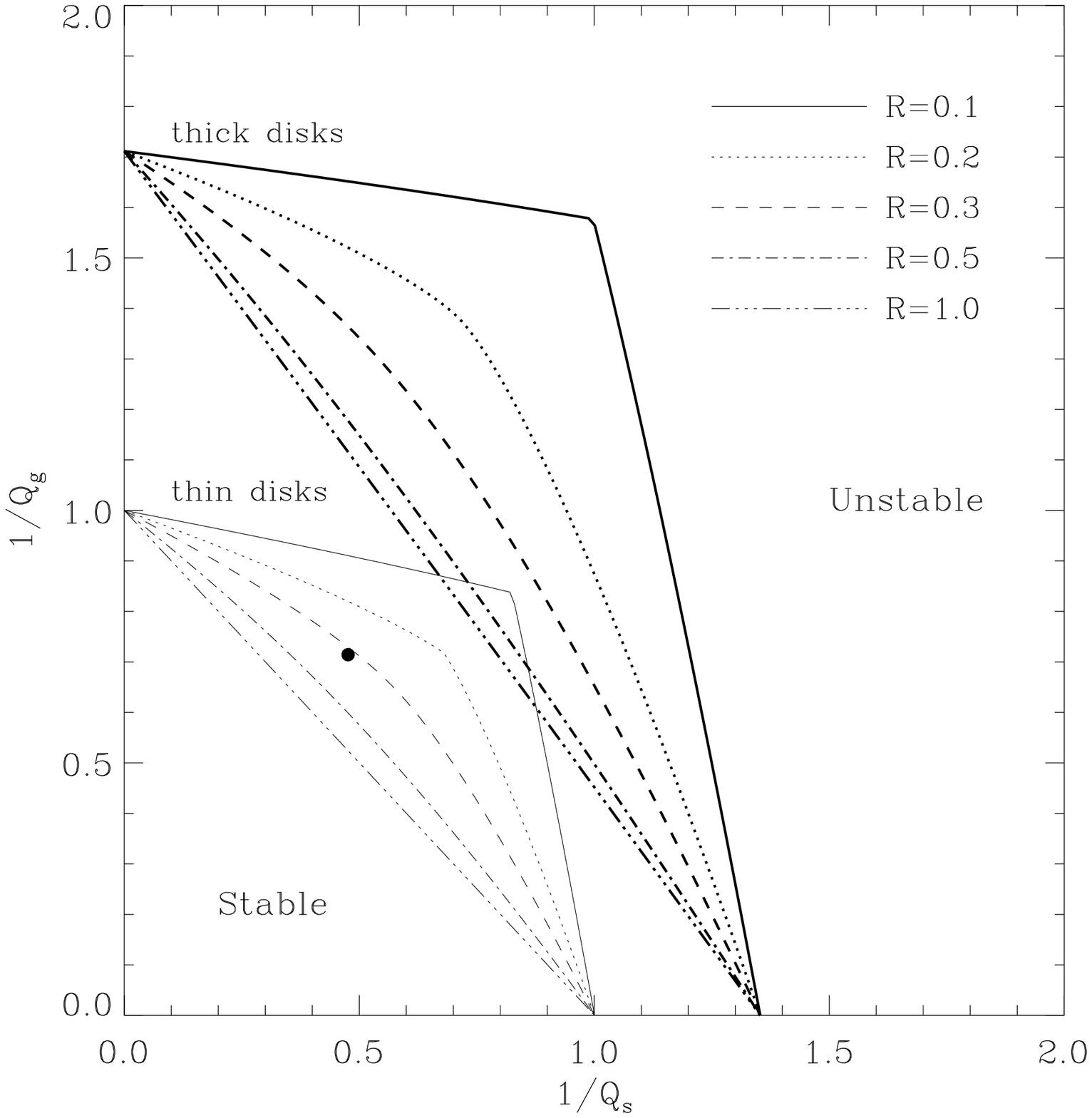}
\caption{
Marginal stability curves in the $Q_s$--$Q_g$ plane for 
axisymmetric gravitational instability in two-component disks 
with $R\equiv c_g/\sigma_x$. 
The heavy curves correspond to thick disks with $H_g=170$ pc and 
$H_s=330$ pc, while the light curves are for razor-thin disks.
The regions below each marginal curve represent conditions for which 
two-component disks are axisymmetrically stable.  The solid circle at
$(Q_s,Q_g)=(2.1,1.4)$ marks solar neighborhood parameter values which
are very stable when the disk is treated as thick,
but would be barely stable if the disk were assumed razor-thin. 
\label{marg}}
\end{figure}

We determine the marginal stability
condition by solving equation (\ref{disp}) with $\omega^2=0$, and plot
the results in Figure \ref{marg} for selected values of $R\equiv
c_g/\sigma_{s,x}$.  
Thick curves are for thick disks with scale heights $H_g=170$ pc and $H_s=330$ pc.
Results from \citet{raf01} for razor-thin disks are also plotted as
light curves for comparison.  The parameter domain below each marginal
curve corresponds to the stable regime.  It is apparent that
axisymmetric gravitational instability is mainly determined by the gaseous
component for small $R$ and $Q_s\simgt 1$.  When $R$ is small, the
characteristic unstable length scales in (standalone) gaseous disks
are much smaller than those of the stellar disk.  In this case, the
relative dominance of the gaseous to stellar components changes rather
abruptly as $Q_g$ and $Q_s$ vary along a given marginal stability
curve, as manifested by the cusp in the $R=0.1$ curve.  As $R$
increases, the distinction between the characteristic length scales in
the gas and stars becomes smaller, and the marginal curve bends
smoothly.  A system with solar neighborhood parameters, marked by the
dot at $(Q_s,Q_g)=(2.1,1.4)$, is highly stable to axisymmetric modes
when treated as thick, while it would be marginally stable under the
thin-disk approximation.

\begin{figure}
\epsscale{1.00}
\plotone{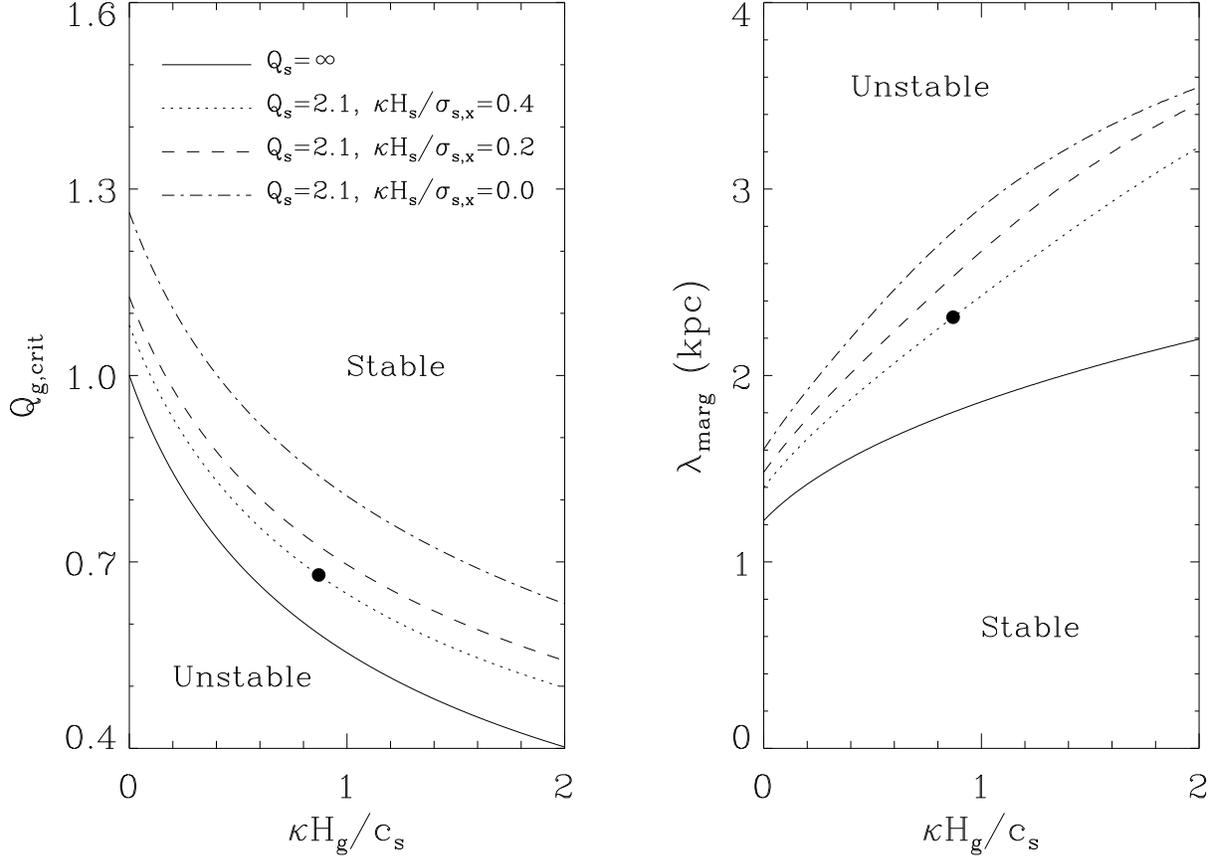}
\caption{
({\it Left}) Critical $Q_g$ values and ({\it right}) the marginal 
perturbation wavelength $\lmarg$ as functions of the gaseous scale 
height $H_g$.  Regions with $Q_g<\Qgcrit$ or $\lambda >\lmarg$ 
correspond to a unstable parameter set.
The solid lines correspond to the gas-only systems,
while other lines are for combined disks with $Q_s=2.1$.  The dots 
in both panels represent the solution for solar neighborhood parameters.
\label{Qcrit}}
\end{figure}

For varying disk thicknesses, Figure \ref{Qcrit} plots the
$\Qgcrit$ values, and the corresponding
marginal wavelength  of perturbations, $\lmarg$ ($\lambda<\lmarg$ modes
are stable).
Self-gravity weakens when $H_g$, $H_s$, or $Q_s$
increases, tending to require lower $\Qgcrit$.  A {\it razor-thin}
gas-only disk has $\Qgcrit=1.0$.  This value increases to
$\Qgcrit=1.27$ if the contribution from a {\it thin} stellar component
with $Q_s=2.1$ is considered.  For disks with thicknesses
$H_g=0.87c_g/\kappa$ and $H_s=0.4\sigma_{s,x}/\kappa$ with $Q_s=2.1$
(similar to the solar neighborhood conditions), however, we have
$\Qgcrit=0.67$ at $\lmarg \sim 2.3$ kpc, as marked by dots in Figure
\ref{Qcrit}.  This indicates that the stabilization caused by finite
disk thickness is considerable, even larger than the destabilizing
effect of stars. Solutions of equation (\ref{disp}) show
that when $H_g=0.87c_g/\kappa$ and $H_s=0.4\sigma_{s,x}/\kappa$ are fixed,
thick disks with $Q_g>1.4$ are stable to axisymmetric
perturbations unless $Q_s<0.8$ (which is a very unrealistic range).
Therefore, the naive expectation that including a stellar component
increases $\Qgcrit$ over unity fails in disks with realistic
thickness.  The fact that $\Qgth \sim 1.4$ for observed star formation
thresholds in disk galaxies has often been attributed to gravitational
instabilities in two-component systems (e.g., \citealt{ken97}).
However, our results show that {\it axisymmetric} modes cannot be providing
the needed instability, when disk thickness is properly considered.
We will show in \S5 that instead, swing amplification of 
{\it non-axisymmetric} perturbations are in fact able to drive
two-component thick disks with $Q_g\simlt 1.4$ into eventual
gravitational runaway.

\section{Numerical Method and Code Tests}

\subsection{Numerical Method}

We have run a number of two-dimensional simulations for star plus gas systems,
varying $Q_g$ and the number of stellar particles while fixing
$H_g$, $H_s$, and $Q_s$.
All the models have $256\times 256$ resolution.
We follow the nonlinear evolution of the gaseous part using the 
same version of the ZEUS code \citep{sto92} as in Paper I;  
the reader is refereed to Paper I for a detailed description 
of the code and the boundary conditions we adopt.
In this subsection, we describe the numerical methods mainly for
the stellar part.   

We use up to $\Nptl=2\times 10^6$ particles to represent a stellar disk
and evolve them based on a particle-mesh (PM) algorithm. 
With its high velocity dispersion, the stellar component has
a characteristic Jeans scale much larger than that of gas,
so the resolution limit imposed by the PM
method is entirely tolerable.  In addition, the
relaxation time of the particles amounts to 
$ t_R = \sigma_{s,x}^3 \Delta x /(\pi G^2 \Sigma_{s0} m)
\sim 1 \times 10^3 \;\torb $ for the parameters we adopt,
where $\Delta x$ is the grid size and $m$ is the mass of individual 
particles \citep{ryb71}.  This time is long enough to ensure 
that the evolution of the stellar system in our PM code is not seriously 
influenced by particle noise and relaxation (e.g., \citealt{whi88}).

The initial distribution (eq.\ [\ref{dist}]) of the particle positions and
velocities are realized using a pseudo-random number generator.
At each time step, the stellar surface density $\Sigma_s$ is calculated on a
$256\times 256$ mesh via the triangular-shaped-cloud 
assignment scheme \citep{hoc88}.   We solve equation (\ref{thick}) by 
employing the fast Fourier transform method on a sheared coordinate grid
in which the surface density is perfectly periodic \citep{gam01}.
The calculated potential is differentiated to give the gravitational
force field at Cartesian mesh zone centers, which is then interpolated back to the
particle locations.  We integrate equation (\ref{eom}) using
a modified predictor-corrector scheme which is second order  accurate in time
(e.g., \citealt{mon89}).  
To handle the kinematics of the background shear self-consistently,
we apply the shearing-box boundary conditions in which $y$-boundaries 
are perfectly periodic and the $x$-boundaries are shearing-periodic
\citep{haw95,hub01}.  These boundary conditions are fully consistent with 
our local models; particles leaving the simulation box from 
one $x$-face re-enter with shifted $y$ positions though the opposite $x$-face. 

\subsection{Code Tests}\label{test}

We have checked our numerical methods on a number of test problems:
one-dimensional or two-dimensional waves in gaseous or stellar disks 
separately or in combination.  In this subsection we present the
results of three tests that not only verify the accuracy of our numerical
code, but also provide information on the necessary number of stellar 
particles to prevent numerical artifacts in the evolution of 
a self-gravitating stellar system.

\begin{figure}
\epsscale{1.00}
\plotone{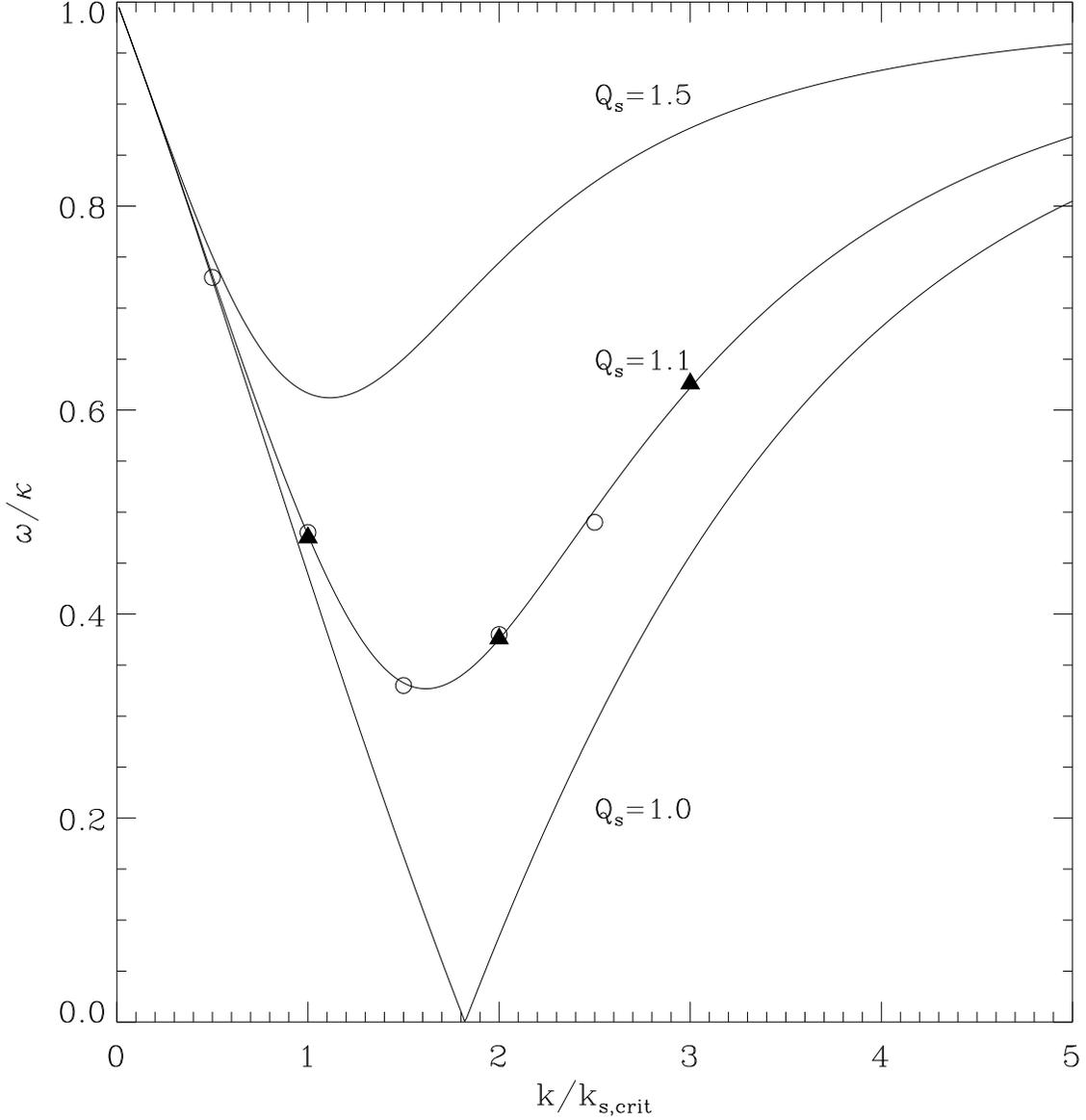}
\caption{
Oscillation frequencies of axisymmetric stellar density waves.
The abscissa is the wavenumber $k$ normalized by 
$\kscrit\equiv \kappa^2/(2\pi G\Sigma_s)$.
One-dimensional, razor-thin, star-only disks with $Q_s=1.1$ are adopted 
and $\Nptl=10^4$ particles are used.
Open circles and solid triangles are from the $X=2$ and $X=1$ models, 
respectively, both of which are in good agreement with the
analytic solutions ({\it lines}) of equation (\ref{disp}).
\label{1d_wave}}
\end{figure}

Figure \ref{1d_wave} plots the test results of axisymmetric density
waves in razor-thin, stellar disks.  For these tests, we consider
one-dimensional, star-only disks with $H_s=0$, $q=1$, $Q_s=1.1$ (close
to the marginal state), and $X=1$ or 2.  For each run, $\Nptl=10^4$
particles are used.  Frequencies obtained from the spatial and
temporal Fourier analyses of the surface densities are given as solid
triangles and open circles for the $X=1$ and 2 cases, respectively.
The numerical results are within $\sim 2.5\%$ of the analytic
predictions (solid lines) from equation (\ref{disp}) for the chosen
set of disk parameters.  This confirms the stellar code's performance.

\begin{figure}
\epsscale{1.00}
\plotone{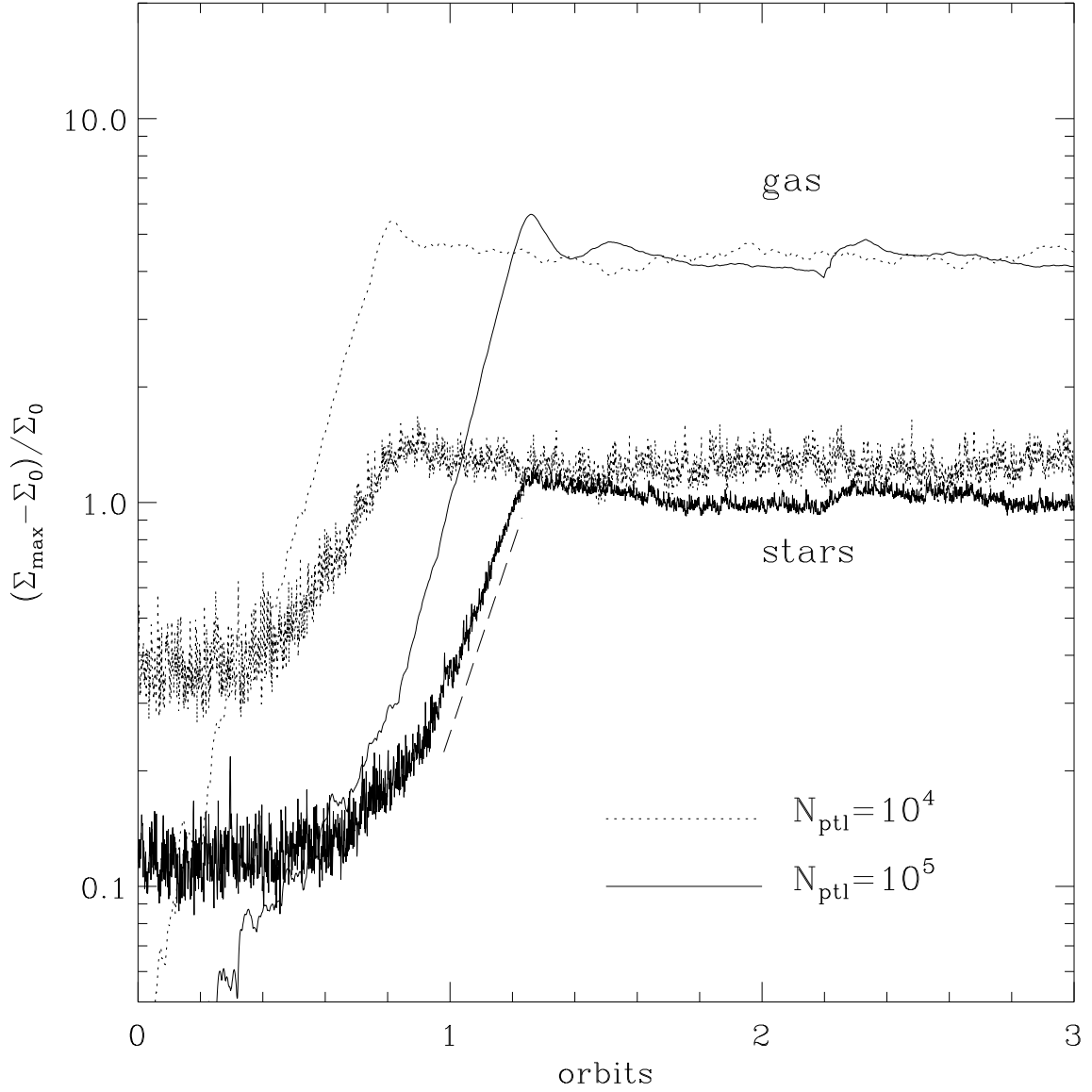}
\caption{
Time evolution of maximum surface densities of gas and stars in
a combined disk with $q=1$, $Q_g=1.2$, $Q_s=1.5$, $H_g=0.05 c_g/\kappa$,
$H_s=0.1\sigma_x/\kappa$, and $c_g/\sigma_x=0.4$.  
An adiabatic equation of state with $\gamma=5/3$ is assumed for the gas.
The long-dashed line gives the theoretical
growth rate of $\sim 0.894\Omega_0$, which is in good agreement with
the results of numerical simulations.
\label{1d_comb}}
\end{figure}

The second test addresses axisymmetric gravitational instability 
of a two-component disk, for which we choose $Q_g=1.2$, $Q_s=1.5$, $q=1$, 
$H_g=0.05c_g/\kappa$, $H_s=0.1\sigma_{s,x}/\kappa$, 
and $c_g/\sigma_{s,x}=0.4$.  
The linear theory in \S\ref{linear} predicts that such a combined disk 
is unstable with a maximum growth rate of $\sim 0.894\Omega_0$, 
which occurs at $\lambda_{\rm max}=3.53\sigma_{s,x}/\kappa$.  Setting 
the one-dimensional 
box size $L=\lambda_{\rm max}$ and assuming the gas is adiabatic with 
index $\gamma=5/3$, we have run two models with 
differing $\Nptl$.  Figure \ref{1d_comb} plots the resulting time 
histories of the maximum surface densities in both gaseous and 
stellar components.  The dotted and solid lines correspond to $\Nptl=10^4$ 
and $10^5$ cases, respectively.
Poisson noise in the stellar distribution immediately 
imposes (via gravity) 
strong density perturbations in the gaseous disk, which would
otherwise be uniform.  The perturbations in stars and gas soon become
coherent and grow exponentially.  The model with $\Nptl=10^4$ has larger Poisson noise,
hence unstable modes grow sooner.  Note that the growth rate of 
stellar surface density in both models is in good agreement with the 
theoretical estimate for the fastest-growing mode, as represented by the long-dashed line 
in Figure \ref{1d_comb}.  The gaseous component, already driven
nonlinear, has slightly faster growth.

Finally we report on the results of an important two-dimensional test.
Evolution of linear waves in two-dimensional, shearing disks in
principle would provide stringent tests for the current purposes, but
we are unaware of test problems that have analytic solutions for
comparison with numerical results.  Swing amplification in a combined
disk would also be a useful test, but it is difficult to accurately
measure the amplification magnitude from numerical simulations.
Propagating non-axisymmetric waves in a shearing disk also would be an
excellent test problem for the gas component (cf. Paper I), but
because of the Lagrangian character it does not work as well for a
stellar disk.  We thus consider evolution of non-axisymmetric
perturbations in non-shearing disks, for which equation (\ref{disp})
still holds true if $q=0$ and $k=(k_x^2+k_y^2)^{1/2}$ are used.  We
have run a set of numerical simulations for stable/unstable waves in
various types of non-shearing disks, and confirmed that numerical
results are consistent with the analytic predictions, verifying again
the accuracy of our numerical code.

\begin{figure}
\epsscale{1.00}
\plotone{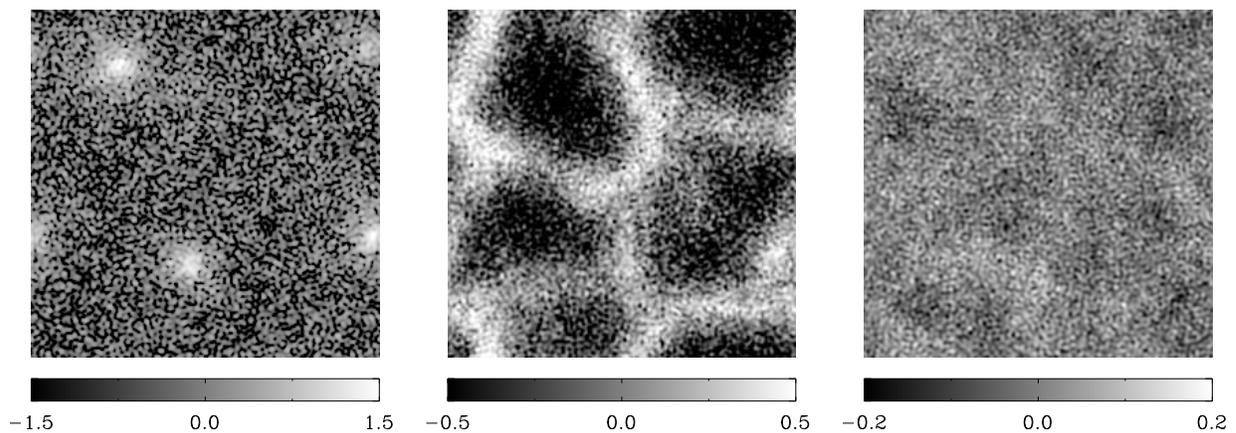}
\caption{Snapshots of surface density at $t/\torb=7.0$ 
in non-shearing, star-only systems
with $Q_s=1.2$, $X=2$, $H_s=0$, but with differing particle 
numbers:  ({\it left}) $\Nptl=10^4$; ({\it middle}) $\Nptl=10^5$; 
({\it right}) $\Nptl=10^6$.  
Gray-scale bars give $\log (\Sigma_s/\Sigma_{s0})$.
It is apparent that large Poisson noise associated with  small $\Nptl$
effectively increases the surface density locally, lowering $Q_s$ and thus
destabilizing systems that would otherwise remain stable.
\label{2d_test}}
\end{figure}

Of particular interest is the effect of Poisson noise on gravitational
instability in a stellar (or combined) system near the marginal state.
Figure \ref{2d_test} plots the snapshots of surface density (in 
logarithmic scale) at
$t/\torb=7.0$ in a razor-thin, rigidly-rotating, star-only disk with
$q=0$, $Q_s=1.2$, and $X=2$.  Although the linear theory suggests that
a disk with these parameters should be stable, the maximum surface
density in a model with $\Nptl=10^4$ grows secularly, prompting the
formation of 4 clumps, as the left panel of Figure \ref{2d_test}
shows.  The growth of perturbations in a model with $\Nptl=10^5$ is
slower, displaying over-density honeycomb structures (middle panel),
while a model with $\Nptl=10^6$ has a stable density field oscillating
mildly (right panel).  The mass-assignment scheme in our
two-dimensional PM code yields Poisson noise at the level $\delta
\Sigma_{s0}/\Sigma_{s0} =1.44 (\Nptl/10^4)^{-1/2}$, as measured by the
standard deviation $\delta\Sigma_{s0}$ of the initial surface density
$\Sigma_{s0}$.  Some fraction of this noise may add to $\Sigma_{s0}$,
decreasing the effective value of $Q_s$.  In cases when $\Nptl$ is
small and when $Q_s$ is slightly larger than unity, $Q_{s,\rm eff}$
can be smaller than the critical value and perturbations may grow
artificially\footnote{ If a half of $\delta\Sigma_{s0}$ contributes
  directly to $\Sigma_{s0}$, the effective Toomre parameter becomes
  $Q_{s,\rm eff}=Q_s (1+ \delta \Sigma_{s0}/2\Sigma_{s0})^{-1}$.  When
  $Q_s=1.2$, $Q_{s,\rm eff}=0.70$, 0.98, and 1.12 for $\Nptl=10^4,
  10^5$, and $10^6$, respectively, which is roughly consistent with
  the results of our numerical experiments.}.  This implies that
Poisson noise can erroneously alter the fate of a numerically-modeled
stellar system, especially when it is close to being marginal.

This point has sometimes been overlooked in previous $N$-body
simulations of disks. 
For example, \citet{gri99} simulated dynamics 
of non-shearing, local stellar disks using about 6,500 particles
and found that disks with $1<Q_{s0}\simlt1.5$ were unstable.
They argued that the apparent discrepancy between the numerical and
theoretical results might be partly because \citet{lin66}'s
WKB theory for density waves is incomplete and partly because 
their method for computing self-gravity is not very accurate.  
We suggest that large Poisson noise alone,
simply associated with a small number of particles, can explain
their numerical results.  Adequate resolution is particularly an issue for 
global disk models, since the Poisson noise on a given spatial scale
is proportional to the linear size of the disk.  Some of the global
simulations of \citet{li05} and \citet{li06}  (with $0.2-1.2\times 10^6$ star
particles for the whole disk), for example, may be affected by
amplification of noise.

\section{Nonlinear Simulations}

To investigate nonlinear evolution of two-component
differentially-rotating disks with a range of gas mass fractions, 
we have run a number of non-axisymmetric
models.  We fix $q=1$, $Q_s=2.1$, $X=2$, $\kappa H_s/\sigma_{s,x}=0.4$, 
$\kappa H_g/c_g=0.87$, and vary $Q_g$ in the range of 1 to 3.  For
these parameters, axisymmetric perturbations are stable.
Solutions for self-consistent vertical equilibria show that
$H_g$ decreases only by $\sim20\%$ as $Q_g$ decreases from 1.4 to 1.0
\citep{kos02}, so that we expect using a fixed value of $H_g$ does not
significantly impact our quantitative results.

We set up initially  uniform gaseous and stellar disks, and apply 
density perturbations only to the gaseous component. Due to randomness
in particle locations, 
the stellar part already has white-noise density perturbations with 
amplitudes of $\sim 10\%$ when $\Nptl=2\times 10^6$ 
which will be immediately transmitted into the gaseous disk. 
In order to keep the Poisson noise from
dominating the evolution of the gaseous disk,  we impose rather
strong gaseous density perturbations using a Gaussian random field 
with a power spectrum $\langle|\Sigma_{g,k}^2|\rangle \propto k^{-8/3}$ 
for $1\leq kL/(2\pi)\leq256$ in Fourier space.  This 
corresponds to a two-dimensional Kolmogorov spectrum if
wave motions obey a sonic dispersion relation (Paper I).   
We measure the standard deviation of the gas surface density in real space 
and fix it to $10\%$.  
Although this level of density fluctuations is chosen as a compromise 
between maintaining the initial perturbations in the linear 
regime and reducing the effect of Poisson noise on 
the simulation outcomes, it may in fact well represent initial conditions 
for the highly-turbulent ISM observed in real disk galaxies. 

\subsection{Threshold for Gravitational Runaway}\label{nswing}

\begin{figure}
\epsscale{1.00}
\plotone{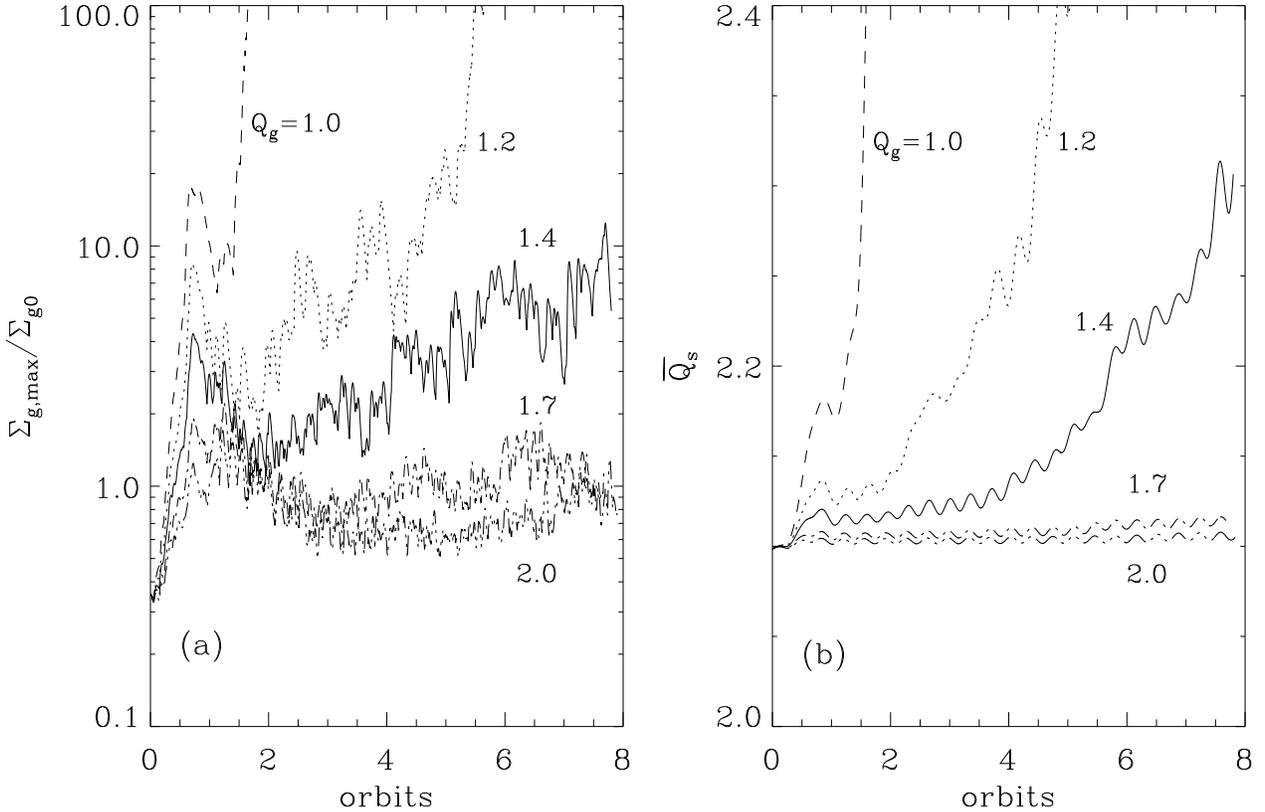}
\caption{Evolution of ({\it left}) the maximum gaseous surface density
and ({\it right}) the spatially-averaged stellar Toomre parameter
$\bar Q_s$, for models with various $Q_g$.  
All the models are two-dimensional and employ $2\times 10^6$ stellar
particles with $Q_{s,\rm init}=2.1$. 
Models with $Q_g\leq1.2$ 
experience gravitational runaway, while models with $Q_g\geq1.7$ 
are stable; the $Q_g=1.4$ model is marginal with a rapidly fluctuating
density field.  For unstable or marginal models, $\bar Q_s$
increases considerably over time due to the gravitational interactions of
stellar particles with both gaseous and stellar concentrations.
\label{2d_evol}}
\end{figure}

\begin{figure}
\epsscale{1.00}
\plotone{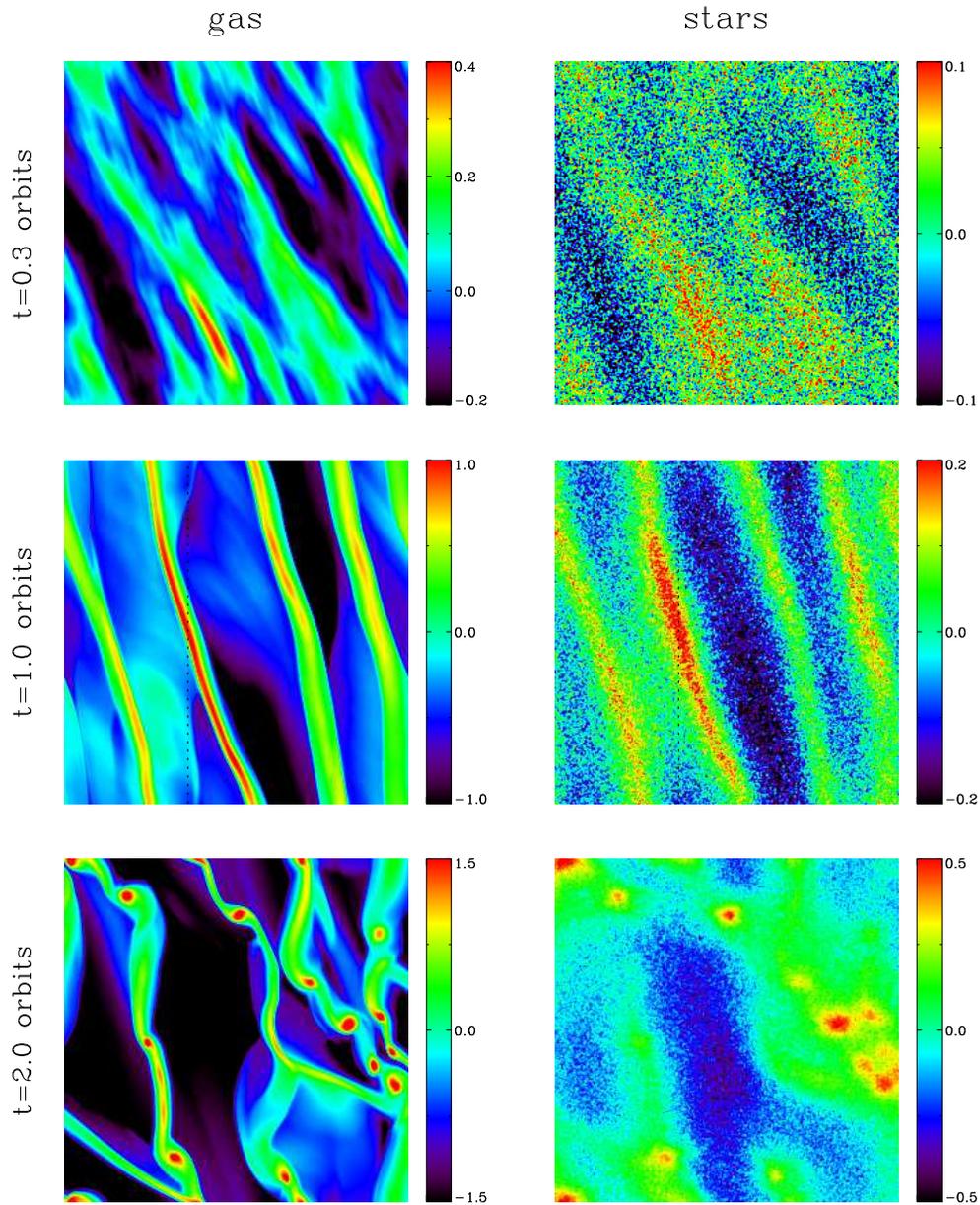}
\caption{Snapshots of ({\it left}) gaseous and ({\it right})
stellar density in logarithmic color scale
for the $Q_g=1.0$ model.   Dotted lines in black in the middle row
indicate the $x$-position at which slices of various physical quantities
are taken in Figure \ref{spiral}.  Strong filaments develop in the
gas, and weaker filaments in the stars, due to swing amplification
($t=1.0$ orbits).  The gaseous filaments subsequently gravitationally 
fragment into clumps ($t=2.0$ orbits).
\label{snapQ10}}
\end{figure}

\begin{figure}
\epsscale{1.00}
\plotone{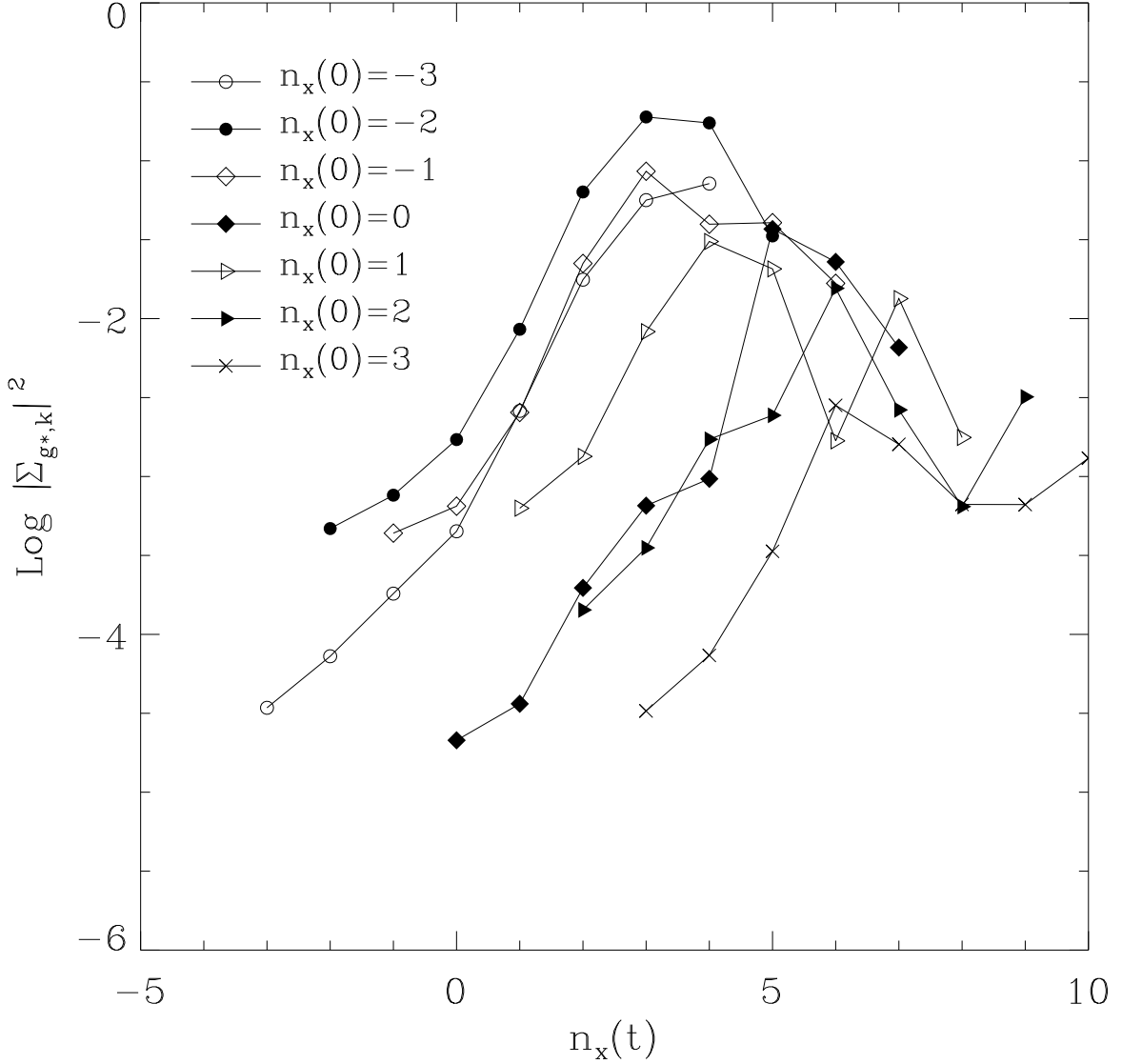}
\caption{Modal growth of power spectra $|\Sigma_{g,k}|^2$ of the gaseous 
surface density against the normalized wavenumber 
$n_x(t)\equiv L k_x(t)/(2\pi)$ 
in the $Q_g=1.0$ model for $t/\torb \simlt 1.1$.  A few selected modes 
with $k_y = 2\pi/L$ and $|n_x(0)|\leq3$ are shown.  
These loosely wound waves exhibit growth via swing amplification, then
saturation.
Note that $n_x(t)$ increases linearly with time, with
$\Delta t= \Delta n_x(t) \torb/(2\pi)$.  The $t=0$ point is the
farthest left on each curve.    
\label{modes}}
\end{figure}

We begin with a high-resolution model with $Q_g=1.0$ and 
$\Nptl=2\times 10^6$.  For this model, the combined gravity of gas and stars 
induces efficient swing amplification, leading to gravitational
runaway.  Because disk thickness dilutes self-gravity, 
each disk (i.e. purely gaseous or purely stellar) in isolation would have remained stable both to axisymmetric
and non-axisymmetric perturbations.
Figure \ref{2d_evol} plots as dashed lines the evolution 
of the maximum gaseous surface density and the spatially-averaged,
stellar Toomre parameter ${\bar Q_s}$ of this model (together with
those of other models). 
Figure \ref{snapQ10} displays snapshots of gas and stellar 
surface density in this model at three time epochs, 
while Figure \ref{modes} shows the initial modal growth of perturbations 
in the gas surface density over time.

Initially, density perturbations (a superposition of modes with
different wavenumbers) adjust by launching sound waves in the gas and
redistributing stellar particles on their epicyclic orbits.  During
this relaxation phase, which lasts $\sim0.2\torb$, coherent
perturbations in stars and gas begin to develop through
gravitational interactions.  Due to the uniform background shear,
the radial wavenumber $k_x(t)$ of perturbations increases linearly with 
time as
\begin{equation}\label{kx}
k_x (t) = k_x(0) + q\Omega_0 k_y t,
\end{equation}
where $k_y$ denotes the $y$-wavenumber and $k_x(0)$ is the initial
$x$-wavenumber of perturbations.  This kinematic shift of $k_x(t)$ 
causes wavefronts to swing from leading to trailing.  
Epicyclic motion of both stellar particles and gaseous fluid elements 
has the same
rotational sense as the wavefronts, extending their exposure to
self-gravity as they linger in wave crests.  Perturbations keep
growing until the radial wavenumber becomes large, forming over-dense,
trailing wavelets (\citealt{gol65b,jul66,too81}; Paper I).  

The modal growth of density perturbations via swing amplification in the
$Q_g=1.0$ model is well
illustrated in Figure \ref{modes} where we plot the evolution of
power spectra of the gaseous surface density as functions of 
the normalized wavenumber $n_x(t) \equiv L k_x(t)/(2\pi)$.
Only the modes with $k_y=2\pi/L$ and $|n_x(0)| \leq 3$ that
have large amplitudes are shown. 
Note that a unity increment of $n_x(t)$ along each curve corresponds to 
a time elapse of $\Delta t = (2\pi)^{-1}\torb$. 
It is apparent that swing amplification is active only when
waves are loosely wound with $|n_x(t)| \simlt 3-4$ above which sonic 
oscillations quench the perturbation growth.
The $n_x(0)=1$ mode that happens to have the largest initial power
grows and emerges first, as manifested by the trailing modes with
$n_x(t/\torb=0.3) \approx 3$ shown in the top panels of Figure 
\ref{snapQ10}.  However, it is the modes with $n_x(0)<0$ that 
eventually dominate the initial swing amplification,
since they have longer time for growth.

As Figures \ref{2d_evol} and \ref{modes} show, the initial
swing amplification in the $Q_g=1$ model
saturates at around $t/\torb\simeq0.7$.  At this time, the gaseous
surface density reaches $\Sigma_{g, \rm max}/\Sigma_{g0}=18.7$, while
the stellar component reaches only $\Sigma_{s, \rm
  max}/\Sigma_{s0}=2.5$.  The stellar component's density increases
less due to its larger initial values of $Q_s$ and
$H_s$, and the steady heating associated with particle
scattering off shearing wavelets.
While the maximum gas surface density is rather high when the swing 
amplifier saturates, the gas filaments are very thin.
Reduction of self-gravity due to finite disk thickness prevents these
filaments  from immediately undergoing gravitational collapse.  The
filaments expand slightly,
reducing their surface density.  Some filaments have sufficient 
perturbed velocities to offset the tendency
for wave rotation of the background shear.  Consequently, the pitch angles
of these high-density filaments vary little; they can be 
viewed as transient (swing-generated) spiral density waves in a 
two-component disk.

\begin{figure}
\epsscale{1.00}
\plotone{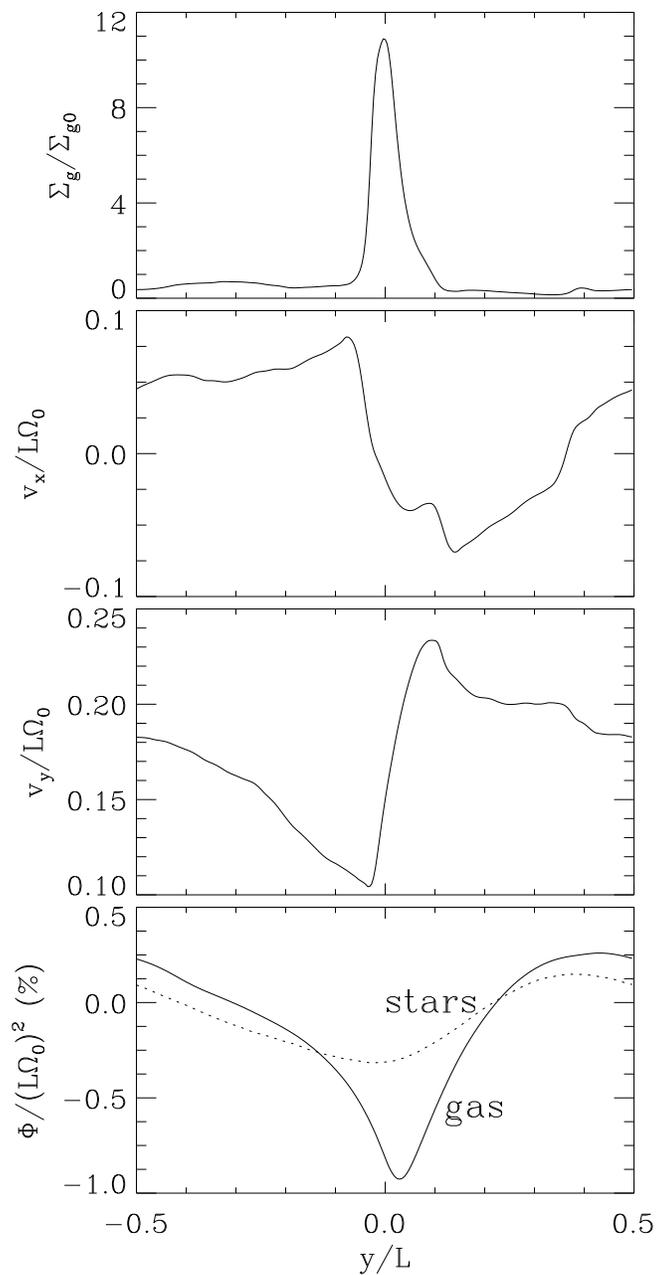}
\caption{Profiles of gaseous surface density, gaseous velocities, and 
gravitational potentials across a filament at $x/L=-0.14$ when 
$t/\torb=1$, for the $Q_g=1.0$ model.  The profiles
are qualitatively similar
to those in galactic spiral shocks when both gas and stars are present 
(e.g., \citealt{lub86}).
\label{spiral}}
\end{figure}

Figure \ref{spiral} plots sample cut profiles of various quantities
along the $x/L=-0.14$ dotted lines in the snapshot shown in the middle
row of Figure \ref{snapQ10}.  The gas filament has a pitch angle
$i=16^{\rm o}$ and is almost corotating with the center of the box; it
coincides with a local stellar enhancement such that the composite may
be considered a transient, local spiral arm segment.  As the gas at
$-0.18<x/L<-0.2$ enters the arm, it is decelerated and compressed,
analogous to a supersonic de Laval flow.  This in turn increases the
velocity parallel to the arm due to the constraint of potential
vorticity conservation (e.g., \citealt{hun64,bal85,gam96,kos02}).
Note that the gravitational potential is dominated by the gaseous
component, which tends to symmetrize the density profile with respect to 
the peak \citep{lub86}.  
The structure of this filament persists
until $t/\torb\sim1.2$ when collisions with other filaments make it
highly self-gravitating.

Paper I found that formation of bound clouds in swing-amplified disks 
occurs generically through one of the three secondary instabilities. 
In order of decreasing self-gravity, these are:
(i) parallel fragmentation of filaments; 
(ii) gravitational collisions of shearing wavelets; and
(iii) rejuvenated swing amplification preceded by nonlinear wave
interactions. 
As explained above, high-density filaments in the $Q_g=1.0$ model
do not immediately fragment due to thick-disk gravity.
Only after several collisions with neighboring filaments is the 
density high enough to form gravitationally bound condensations.
As Figure \ref{snapQ10} shows,  the $Q_g=1.0$ model forms 14 bound 
gaseous clouds, the average mass of which is $\sim6\times 10^7\Msun$,
roughly 10 times larger than the Jeans mass 
$M_{\rm J,2D}=c_g^4/(G^2\Sigma_{g0})$ in the corresponding
gas-only, razor-thin disk.  For comparison, in the $Q_g=1$ models of
Paper I, the condensations that formed had 
$M/ M_{\rm J,2D}\sim 0.5-1$, suggesting that the effect of 
finite disk thickness (combined with an active stellar disk) is 
non-negligible in setting the cloud masses that form.

We note that the bound condensations that form in our models
may represent \ion{H}{1} superclouds observed in many disk galaxies 
(e.g., \citealt{elm83,kna93}).
In our models (isothermal, without feedback), these superclouds keep 
collapsing in a runaway fashion to the resolution limit.
An artificial consequence of the runaway cloud collapse,
at highly nonlinear stages,
is the formation of loose stellar aggregates, as Figure \ref{snapQ10} displays.
Note that all the stellar clumps coincide with the massive gaseous clumps.
In realistic situations, however, gaseous superclouds would presumably fragment 
into lower-mass GMCs when appropriate physical processes such as
internal turbulence are included.
If $M_{\rm sc}$ and $R_{\rm sc}$ denote the mass and size of superclouds, 
respectively, stars would be drawn in strongly only 
if $R_{\rm sc} < G M_{\rm sc}/\sigma_{s,x}^2 \sim 300$ pc 
for $M_{\rm sc}=6\times 10^7\Msun$ and $\sigma_{s,x}=30\kms$.  This
corresponds to a supercloud surface density
$\Sigma_{\rm sc} > 200 \Msun$ pc$^{-2}$, comparable to the 
typical surface density of a GMC.  As long as GMCs are destroyed
by star formation before the whole supercloud reaches a density
as large as that of an individual GMC (which is observationally true),
stellar clumps would not be produced in reality.

As shown in the right panel of Figure \ref{2d_evol}, the stellar velocity 
dispersion in this model  continually  increases as individual stars scatter 
off stellar and, more importantly, gaseous filaments formed via swing 
amplification.  This process is somewhat analogous to heating of a stellar disk
by transient spiral density waves 
(e.g, \citealt{bar67,sel84,car85,fuc01,des04,min06}),
although the runaway fragmentation and collapse of gaseous fragments 
in our two-component models cause the stellar velocity 
dispersions to rise more rapidly than in star-only systems.

Figure \ref{2d_evol} shows that a model with $Q_g=1.2$ also becomes
unstable, but with weaker gaseous gravity it takes longer to reach
gravitational collapse.  The first-generation filaments formed in this
model do not fragment directly, and unlike in the $Q_g=1$ model
collisions do not yield merger-induced fragmentation.  
Instead, the structures nonlinearly interact with each other and supply fresh,
small-$|k_x|$ modes that undergo subsequent swing
amplification (Paper I; see also \citealt{fuc05} for its stellar
analog).  Four successive stages of ``rejuvenated'' swing amplification 
(from interactions of filaments of order unity amplitudes) are
required to drive the $Q_g=1.2$ model into eventual gravitational
runaway.  The average mass of gaseous clouds that form is again 
$\sim 10 M_{\rm J,2D}$.

\begin{figure}
\epsscale{1.00}
\plotone{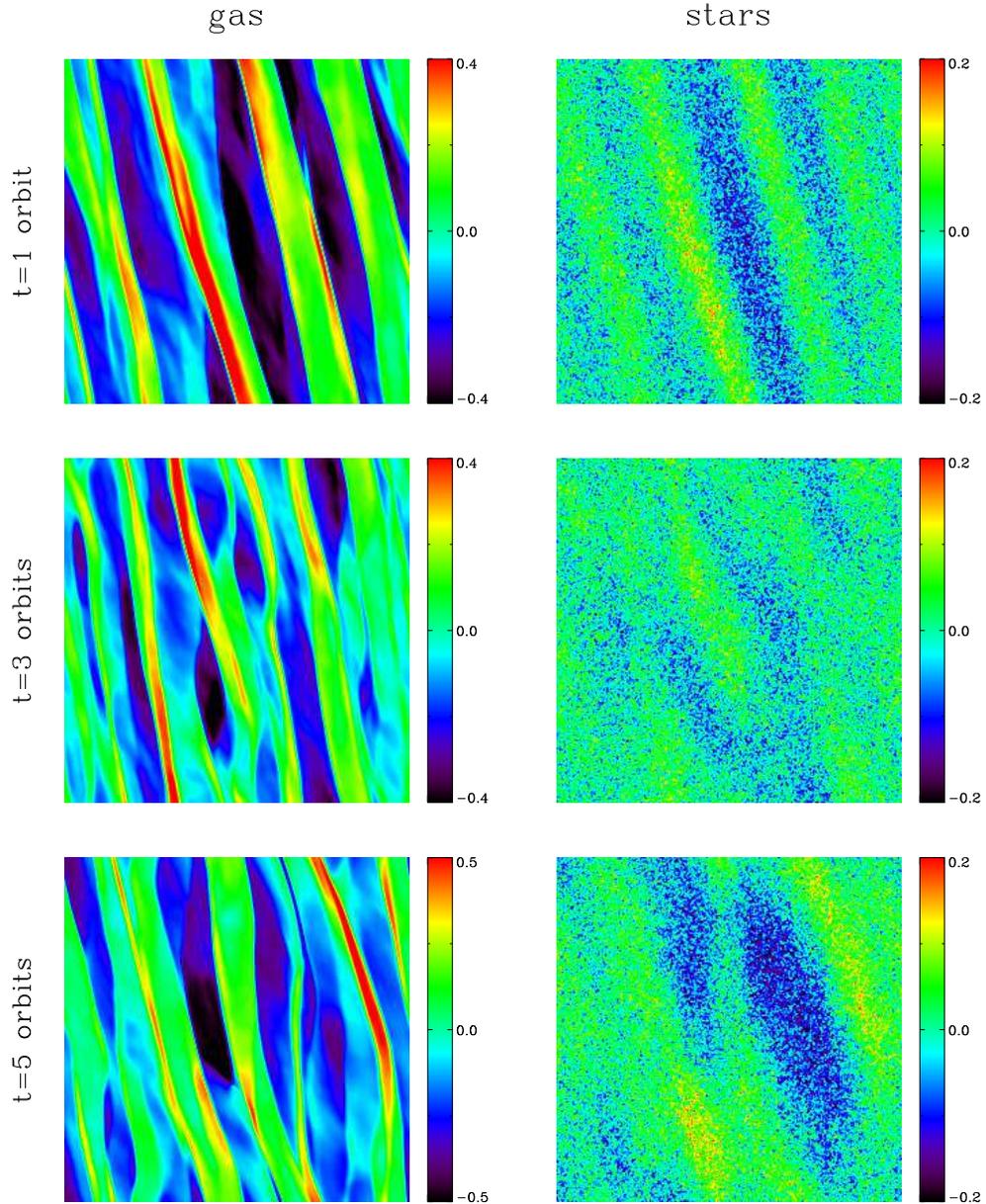}
\caption{Snapshots of ({\it left}) gaseous and ({\it right})
stellar density in logarithmic color scale
for the $Q_g=1.4$ model.  Note that nonlinear shearing wavelets 
are prevalent in the gaseous disk, while density fluctuations in
the stellar disk are very weak.
\label{snapQ14}}
\end{figure}

When we further reduce the gas surface density so that $Q_g=1.4$, the
initial swing amplification yields a moderately self-gravitating state 
that never forms bound clumps over the course of the simulation.
As Figure \ref{snapQ14} shows, 
the gaseous component in this model is dominated by shearing 
wavelets (and the related velocity field).   Mild
rejuvenated swing amplification allows the peak gaseous 
surface density and velocities to grow steadily, reaching the 
nonlinear regime. 
The stellar velocity dispersions also grow, but the net increase is 
less than 10\% at $t/\torb=8$ (see Fig.\ \ref{2d_evol});
the stellar disk also remains virtually uniform with only very low
fluctuations in
density.  We regard this model as ``marginal'' because 
it would almost certainly end up with bound
condensations if evolved over a sufficiently long time.  In other
models with $Q_g\geq1.7$, growth is so weak that
perturbed surface densities remain in the linear regime at saturation
of the first swing amplification. Rejuvenation of swing 
is almost absent in these models, so that the combined systems can be
considered nonlinearly stable.  We conclude, therefore, that the
threshold for gravitational instability in combined gas-star disks has
$Q_g$ near 1.4.

\subsection{Gravity-Driven Turbulence}\label{gdturb}

We now examine the properties of turbulence generated by swing
amplification in two-component disks.  We define $\dEk\equiv
\case{1}{2}\int\Sigma_g({\mathbf v}-{\mathbf v_0})^2
dxdy/\Sigma_{g0}$, representing the specific kinetic energy associated
with turbulent velocities in the the gas.\footnote{Note that the total
  perturbed kinetic energy also contains other terms; $\dEk$ is (half
  of) the mass-weighted mean-squared turbulent velocity dispersion.}  We have shown
in \S\ref{test} that large Poisson noise in a particle distribution
effectively increases the mean surface density of stars in local
regions, and can even incur spurious gravitational instability.
Similarly, random
particle noise  can artificially -- and possibly significantly --
enhance the saturated state value of $\dEk$. 

We want to measure $\dEk$ in a way that minimizes  the effect of
particle noise.  
To this end, we take the following steps:
(i) Run a two-component model in which both stellar and gaseous
disks evolve under the total (gaseous plus stellar) self-gravity, and 
measure the resulting kinetic energy, which we denote by $\dEk^{\rm SG}$.
(ii) Run a control model with the same set of parameters as in step (i),
but evolving the stellar component kinematically 
(i.e., omitting the $\nabla(\Phi_s+\Phi_g)$ gravity terms in 
eq.\ [\ref{eom}]), 
while evolving 
the gaseous component including both self- and stellar gravity.  
In runs of type (ii), no initial perturbation is applied to 
the gas, but the gas responds to the Poisson noise of the stars.  The
energy resulting from runs of type (ii) are denoted 
$\dEk^{\rm NG}$.
(iii) Calculate the gaseous velocity dispersion 
$\sigma_v \equiv (2\langle\dEk^{\rm SG}\rangle
- 2\langle\dEk^{\rm NG}\rangle)^{1/2}$, where
the brackets $\langle \;\rangle$ denote a temporal average over
some time interval (typically over 1-6 orbits).
(iv) Finally, vary $\Nptl$ and repeat steps (i)-(iii) to check
the convergence of $\sigma_v$ with increasing $\Nptl$.

\begin{figure}
\epsscale{1.00}
\plotone{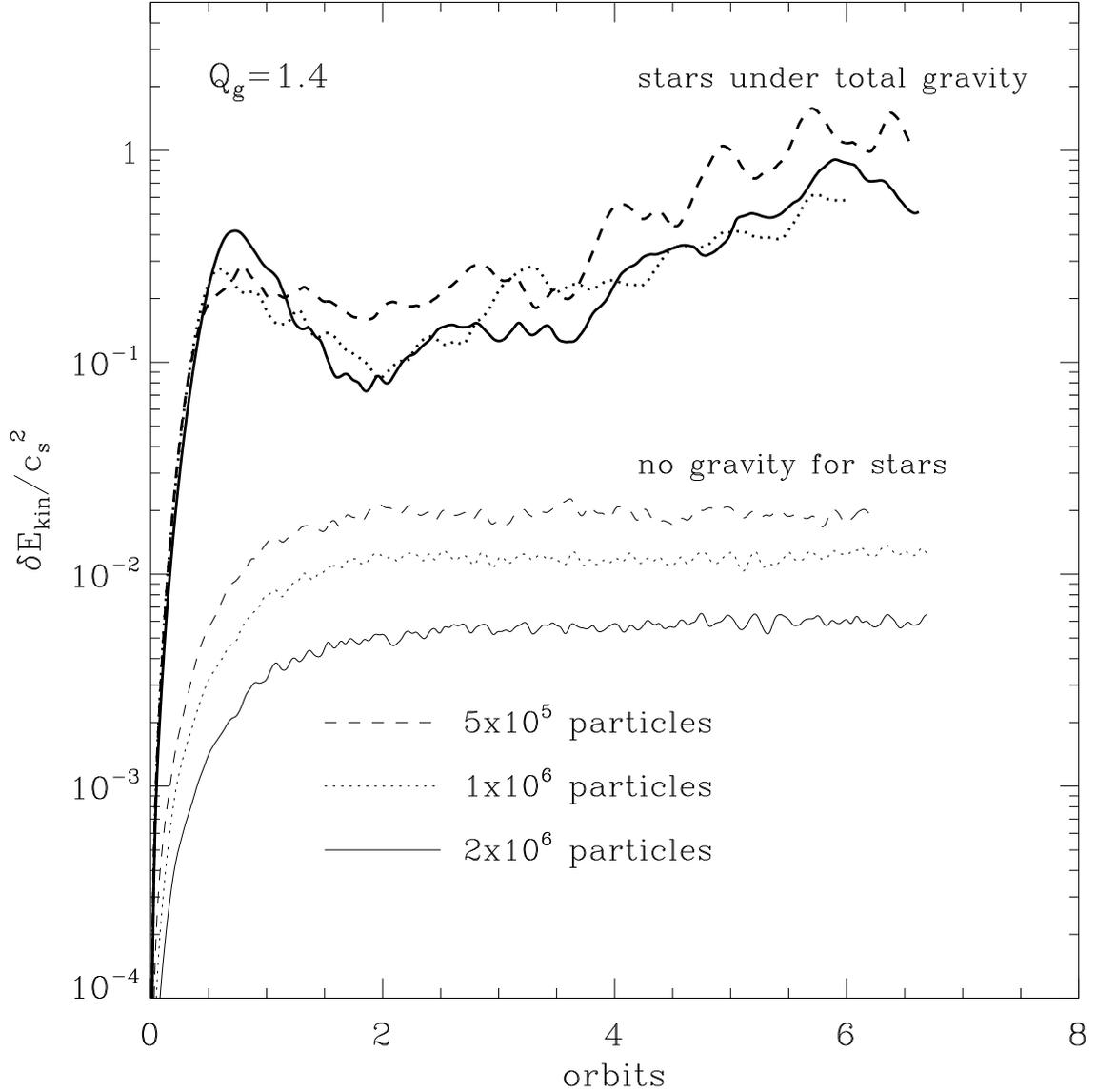}
\caption{Evolutionary histories of the random kinetic energy
of the gas, $\dEk=\case{1}{2}\int \Sigma_g({\mathbf v}-{\mathbf v_0})^2
dxdy/\Sigma_{g0}$, in two-component disks with $Q_g=1.4$ and $Q_s=2.1$.
Three cases with different $\Nptl$ are shown.
The thick curves are for the models where the both stellar and gaseous
components are fully self-gravitating, while the thin lines 
correspond to the cases in which the combined gravity applies only
to the gaseous disk.  The stellar disk in the latter case evolves
passively, but still imposes gravitational forcing from Poisson noise
on the gas disk. 
\label{Q14}}
\end{figure}

Figure \ref{Q14} displays evolutionary histories of $\dEk$ for 
$Q_g=1.4$ models that differ in the number of particles 
and the treatment of gravity for the stellar component.
The thick curves correspond to runs following (i) above, 
while thin lines correspond to runs following (ii).
In the cases with full gravity, the systems relax after the initial 
swing amplification, with $\dEk^{\rm SG}$ decreasing temporarily as some of the gaseous kinetic energy
is transferred via sound waves and/or weak shocks into thermal energy 
that dissipates under the isothermal prescription. 
The continuous input from stellar perturbations as well as nonlinear
feedback from the shearing wavelets cause a resumed increase in
$\dEk^{\rm SG}$.  
After $t=2$ orbits, 
models with gravitating stars show 
a secular increase of $\dEk^{\rm SG}$.
On the other hand, in models where the stars evolve passively, the
initial swing is weak enough that  $\dEk^{\rm NG}$ simply 
saturates at a constant (low) value.
Other models (not shown) with $Q_g\geq 2.0$ 
show that both $\dEk^{\rm SG}$ and $\dEk^{\rm NG}$ are more or less constant 
after $t/\torb=2$, because swing amplification is very mild.

\begin{figure}
\epsscale{1.00}
\plotone{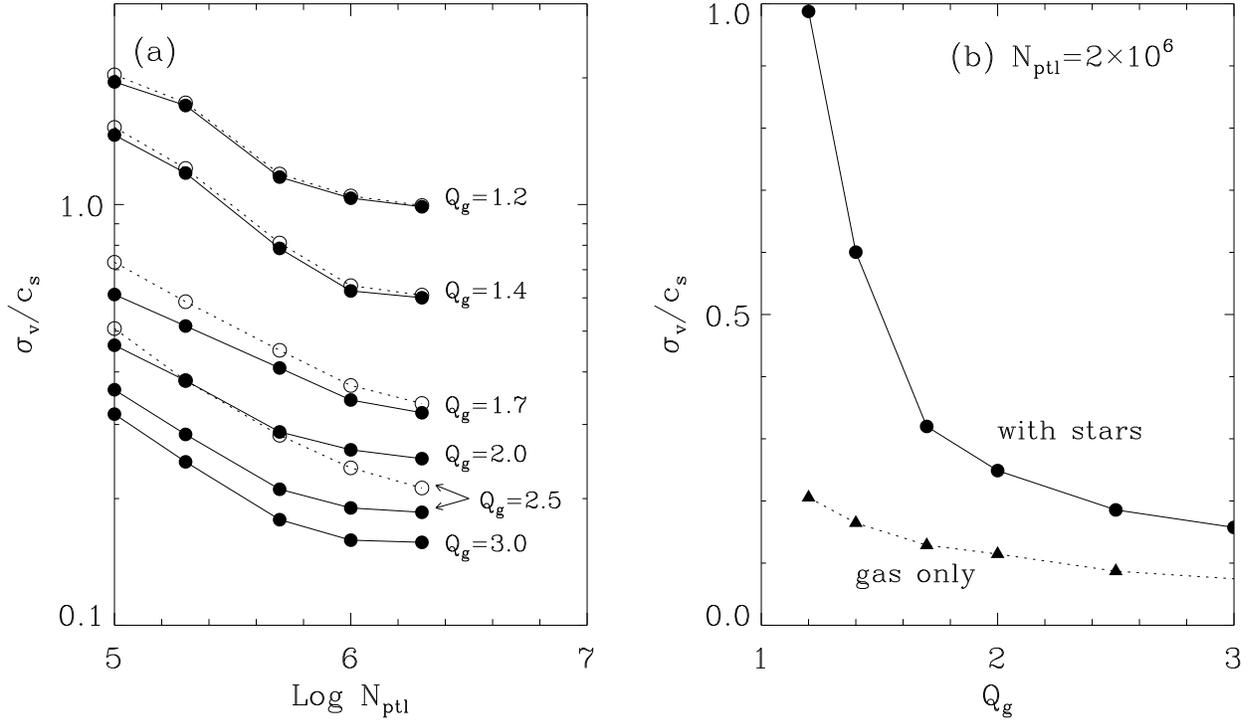}
\caption{Amplitudes of mean fluctuating gaseous velocities $\sigma_v$ 
in two-component models with $Q_s=2.1$, $H_g=0.87 c_g/\kappa$,
and $H_s=0.4\sigma_x/\kappa$, to test numerical resolution effects.
({\it a}) Filled circles show $\sigma_v$ as a function of stellar
particle number $\Nptl$ 
after subtracting the random kinetic energy induced by 
Poisson noise in the stellar distribution, while open circles 
give $\tilde\sigma_v$, the gas velocity dispersion without this 
correction.
({\it b}) Filled circles give $\sigma_v$ for $\Nptl=2\times 10^6$, and 
filled triangles are for gas-only systems.
See text for details.
\label{turb}}
\end{figure}

Figure \ref{turb}(a) plots the results for $\sigma_v$ against $\Nptl$ as filled circles 
for various $Q_g$.  In computing $\sigma_v$, we take time averages
of $\dEk$  over $t/\torb=1-6$ for all the models except the
unstable model with $Q_g=1.2$.  For $Q_g=1.2$, time averages 
are taken only over $t/\torb=1-4$ since afterwards  the velocity field
is mainly a response to discrete high-density filaments.
For comparison, Figure \ref{turb}(a) also plots values of  
$\tilde\sigma_v\equiv \langle 2 \dEk^{\rm SG}\rangle^{1/2}$
as open circles for selected values of $Q_g$.  The figure shows that
for $Q_g<1.7$, $\sigma_v$ and $\tilde\sigma_v$ are in close agreement.
More generally, it is clear that there are progressively smaller
differences between $\sigma_v$ and $\tilde\sigma_v$ as $Q_g$
decreases, indicating that the contribution from swing-amplified 
Poisson noise to 
$\tilde\sigma_v$ becomes increasingly unimportant at lower $Q_g$.

Even after $\dEk^{\rm NG}$ -- embodying the kinetic energy directly 
induced in the {\it gaseous} disk 
by the {\it stellar} Poisson noise -- has been subtracted, Figure
\ref{turb} shows that $\sigma_v$ still 
depends on $\Nptl$.  This is because the gas responds to
swing-amplified perturbations in the stars that partly originate as
noise in the initial stellar distribution. 
Note, however, that $\sigma_v$ clearly converges with increasing
$\Nptl$, implying that this secondary noise effect becomes
negligible.  The convergence study suggests that 
$\sigma_v$ at $\Nptl=2\times 10^6$
provides a clean measure of the turbulent energy that is
uncontaminated by effects of Poisson noise.  Our results also imply 
that numerical simulations of marginally-unstable galactic
disks are not reliable if insufficiently many
particles are employed for the collisionless part.

Figure \ref{turb}(b) plots $\sigma_v$ for two-component disks 
with $\Nptl = 2\times 10^6$ as well as the
velocity dispersion for gas-only disks.  For stable systems with
$Q_g \simgt2.0$, the stellar component enhances the amplitudes of the
gaseous random velocities by about a factor of 2 compared to the
gas-only counterpart, although velocity dispersions are small overall
in this regime.  When the disk is marginal or unstable ($Q_g=1.2$ or
1.4), $\sigma_v$ becomes comparable to the effective speed of sound,
and is $\sim 4-5$ times larger than the value obtained for a gas-only
disk.  This suggests not only that the stellar disk can significantly
affect driving of gaseous turbulence, but also that self-gravity
combined with galactic shear is capable of inducing sonic-level
gaseous turbulence under realistic disk conditions.

\begin{figure}
\epsscale{1.00}
\plotone{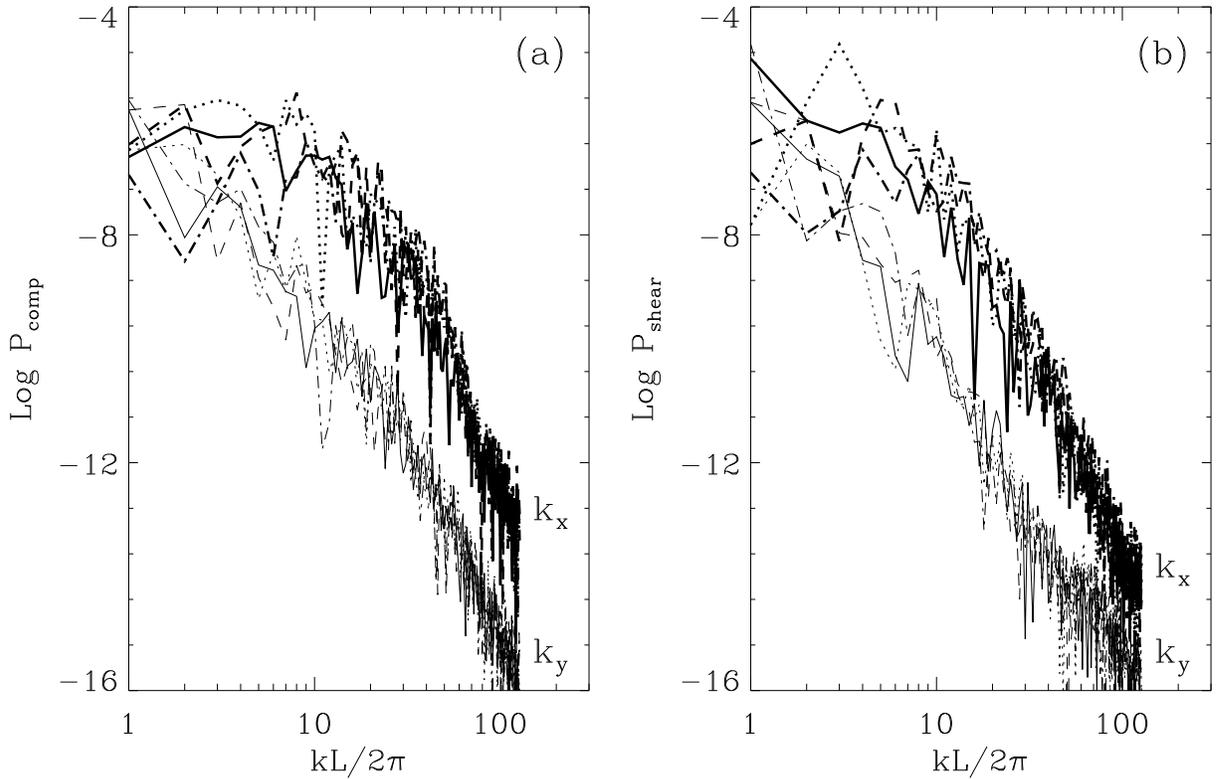}
\caption{Amplitudes of the power spectra along the $k_x$- and $k_y$-axes
of ({\it a}) compressive and ({\it b}) shearing parts of the gaseous
velocity in a two-component model with $Q_g=1.4$ at $t/\torb=2.5$.
Solid line, dotted line, dashed line, and dot-dashed lines 
indicate the modes with $k_yL/(2\pi)=0$, 1, 2, 3 as a function of $k_x$
(heavy lines), and the modes with $k_xL/(2\pi)=0$, 1, 2, 3 as a
function of $k_y$ (light lines), respectively.  The velocity power
is clearly anisotropic and the shearing part dominates at large scale.
Both compressive and shearing parts along the $k_x$-axis show a break near
$k_xL/(2\pi) \sim 8-10$ below which the power flattens.
\label{vpow}}
\end{figure}

To characterize the turbulence generated by 
swing amplification, we plot in Figure \ref{vpow} Fourier power spectra 
of the compressive and shear components of gaseous velocities defined by
\begin{equation}
P_{\rm comp} = |\hat{\mathbf k}\cdot \delta{\mathbf v}_k |^2\;\;\;\;
{\rm and}\;\;\;\;\;
P_{\rm shear} = |\hat{\mathbf k}\times \delta{\mathbf v}_k |^2,\;\;\;\;
\end{equation}
respectively.  Here, $\delta{\mathbf v}_k$ denotes the Fourier 
transform of the perturbed velocity, ${\mathbf v}-{\mathbf v}_0$.
Data at $t/\torb=2.5$ of a fully-gravitating two-component 
model with $Q_g=1.4$ and $\Nptl=2\times10^6$ are used for Figure \ref{vpow}.
We have also calculated the density power spectra and confirmed that 
they follow the compressive velocity power spectra very closely.

The velocity power spectra are clearly anisotropic, as is typical of
turbulence in systems with strong background shear (e.g.,
\citealt{haw95,kim03}).  Comparisons of Figure \ref{vpow}$a$,$b$
reveal that the ratio of total power in the shearing to compressive
parts is about 1.5.  This value is quite small compared to $\sim4-10$
found in recent simulations of hypersonic turbulence (e.g.,
\citealt{bol02,ves03}), presumably reflecting the facts that (i) our
models are subsonic or weakly supersonic at best, so that compressions
are not strongly dissipated, and (ii) driving by self-gravity favors
pumping of compressive motions.  Although the power spectra vary too
much at low $k$ to yield clean power-law indices and are affected by
numerical dissipation at high $k$ (e.g., \citealt{jkim05}), our
results are consistent with $P \propto k_y^{-4.5}$ in the inertial
range.  The cuts along the $k_x$-axis exhibit a break near $k_x
L/(2\pi) \sim 8-10$, below which the power flattens and above which 
$P_{\rm comp} \propto k_x^{-4}$ and $P_{\rm shear} \propto k_x^{-6}$.
About 85\% of the turbulent power is in large scales modes 
with $\lambda \leq L/10$, suggesting that the small-scale turbulent velocity
dispersion is a small fraction of the total; this proportion could, however,
differ in three-dimensional models.
The flattening of power spectra at small $k_x$ and steepening at
large $k_x$ in our models makes
sense, considering that in turbulence driven predominantly by swing 
amplification the background shear tends to increase the $k_x$ value
of any structure over time.  In quasi-steady state, 
nonlinear interactions of shearing wavelets evidently supply fresh power 
into small-$|k_x|$ modes at just the rate needed to compensate for the 
kinematic shift of $k_x$ due to shear, such that  the shape
of the power spectrum is nearly flat.   While turbulent energy
cascades to small scales appear to shape power spectra along the 
$k_y$-direction, shear and nonlinear feedback clearly dominate 
energy flows in the $k_x$-direction.

\section{Summary and Discussion}

\subsection{Summary}

In this paper we have analyzed the nonlinear dynamical evolution of 
two-component galactic disks.  One primary goal has been to extend our
previous work (Paper I), which modeled gas-only disks, to evaluate
thresholds for runaway gravitational instability and formation of
bound clouds.  The other primary goal has been to quantify the
generation of gaseous turbulence in cases where gravitational runaway
does not occur.   The gaseous medium is isothermal and
evolved by a hydrodynamic technique as in Paper I, whereas the 
newly implemented stellar component is represented
by an $N$-body system evolved by a particle-mesh method.  
The two components interact through their respective 
gravitational potentials.  
Our simulations are two-dimensional, but 
incorporate (important) finite disk thickness effects 
in an approximate fashion (see \S\ref{selfg}).  
The local framework  we employ incorporates  galactic rotational 
shear, tidal gravity, and Coriolis forces.  The present 
models do not include effects of
magnetic fields or externally-driven stellar density waves.

Our main results are as follows:

1.  Allowing for nonzero disk thickness significantly stabilizes
axisymmetric modes in two-component disks (see \S\ref{linear}).  For
solar neighborhood conditions with $Q_s=2.1$, the critical Toomre
parameter of the gaseous component would be $\Qgcrit=1.27$ if the both
disks were razor-thin.  Using more realistic scale heights 
$H_g=170$ pc and $H_s=330$ pc for the gaseous and stellar disks,
respectively, the critical value is reduced to $\Qgcrit=0.67$, well
below the observed value.  Two-component, thick disks with $Q_g=1.4$ 
are axisymmetrically 
stable unless $Q_s<0.8$, which is again inconsistent with observed galactic
conditions.  Star formation thresholds at $\Qgth\sim
1.4$ observed in the outer regions of external massive disk galaxies 
(\citealt{ken89}, \citealt{mar01}) are therefore not a consequence of 
{\it axisymmetric} gravitational instability.

2. Our two-dimensional simulations show that two-component disks
undergo gravitational runway when $Q_g< \Qgcrit$ (see \S\ref{nswing}).
For stellar parameters representing the solar neighborhood ($Q_s=2.1$
and $H_s=330$ pc), this nonlinear threshold occurs at $\Qgcrit
\sim 1.4$.  Disks with $Q_g$ below this threshold experience {\it
  nonaxisymmetric} gravitational instability, forming bound
condensations of mass $\sim6\times 10^7\Msun$ each.  These
condensation masses are 10 times larger than those that form in
razor-thin, gas-only disks because larger scales are required for
gravitational instability in thicker disks.  We note that the critical
value of $Q_g$ for non-axisymmetric instability is larger than that for
axisymmetric instability by more than a factor of 2.

3. In addition to forming bound condensations, swing amplification
is able to generate a significant level of gaseous 
turbulence (see \S\ref{gdturb}).  The active stellar component is
important to this, with velocity dispersions a factor $2-5$ larger
than in the equivalent gas-only disks.  
Our simulation results show that when $Q_g=1.2-1.6$, corresponding to 
disks near the stability threshold, 
the density-weighted velocity dispersions of the gas amount to 
$\sigma_v\sim (0.3-1)c_g$.  This suggests that
swing amplification can serve to tap rotational and gravitational
energy in feeding random motions in the ISM.  The 
turbulence in our models is anisotropic and has slightly more energy
in  the shearing than the 
compressive motions. Below $k_xL/(2\pi) \simlt 8-10$, the power
spectrum is relatively flat, while it steepens at larger $k$.
Most ($\sim85\%$) of the total power is contained in large
scale modes with $\lambda \leq L/10$, although in fully three dimensional
models the relative power in small and large scales could potentially
change.

4. Poisson noise in the positions of randomly placed particles
produces initial density fluctuation amplitudes $\delta \Sigma/\Sigma
\approx 1.4 (\Nptl/10^4)^{-1/2}$ for two-dimensional particle-mesh 
simulations with grid resolution $256^2$.  
For near-marginal systems we show
that these density variations can lead to spurious local instabilities
and artificial fragmentation (\S \ref{test}).  Poisson noise can also
lead to overestimates of turbulent velocity fluctuations produced by
swing. Insufficient particle numbers can therefore seriously
compromise the results of $N$-body disk simulations.  Care must be taken
to test numerical convergence.  Here, we find that $\Nptl=1-2\times
10^6$ is needed to avoid contamination by Poisson noise.

\subsection{Discussion}

In this work, we have found that the threshold for formation of
gravitationally bound gas clouds occurs at $\Qgcrit\sim1.4$.  This
result is in fact nearly identical to the nonlinear threshold value
$\Qgcrit=1.3$ we found in Paper I (for unmagnetized models), due to
compensation between two effects that were not accounted
for in Paper I. The current models allow for a (stabilizing) nonzero
disk thickness, and include a (destabilizing) active stellar
component; although each effect by itself makes a significant
difference in $\Qgcrit$, in net they nearly cancel each other.  
While the present models do not include magnetic fields, we expect
that inclusion of magnetic effects would not significantly change the
results; in Paper I, there was only a $10\%$ change in $\Qgcrit$ even
with equipartition-strength large-scale magnetic fields.

An open question is how sensitive $\Qgcrit$ is to the precise
parameters characterizing the stellar disk.  To define a manageable
parameter space, we have fixed the value of $Q_s$ to 2.1, 
the (initial) ratio of the
stellar velocity dispersion to the gas sound speed to 4.3, and the
ratio of stellar to gas 
disk thicknesses to 1.9.  The values chosen for these parameters are based on
solar-neighborhood observations, and thus are representative of
conditions in the middle of the galactic disk for a late-type spiral.  
Although the stellar
component's parameters surely vary from the deep interior to the outer
reaches of a galactic disk, constraining the relevant quantities
observationally is in fact quite challenging.  Extragalactic studies
\citep{van81,van82,deg97} are consistent with the stellar disk
thicknesses being independent of radius at least for late-type
spirals.  In this case, a (self-gravitating) stellar disk with a
constant ratio of vertical to radial velocity dispersion will have
$\sigma_x \propto \Sigma_{s0}^{1/2}$.  For an exponential stellar
surface density $\Sigma_{s0}\propto \exp(-R/R_d)$ with scale length
$R_d$, this would imply that the variation of $Q_s$ over $1<R/R_d<4$
is less than 30\%.  Furthermore, $Q_s$ varies over only a small range
from one galaxy to another; \citet{bot93} found that $Q_s=2-2.5$ fits
the data well (consistent with theoretical predictions of e.g.
\citealt{sel84}).  Thus, although we obtained the result $\Qgcrit\sim
1.4$ for a specific set of stellar parameters, the modest variation in
these parameters among and within real galaxies suggests that there
would be a correspondingly modest variation in $\Qgcrit$.

It is interesting to note that measured values of $Q_g$ at the
observed threshold radii for star formation are around 
$\Qgth\sim1-2$ \citep{ken89,mar01}. 
Unless the stellar properties at these threshold
radii differ dramatically from those we have adopted, the results of
our work, showing $\Qgth$ within the same range, 
quantitatively support the possibility that star formation
thresholds and (nonaxisymmetric) gravitational instability thresholds
are likely to be one and the same.
A similar conclusion was reached very recently
by \citet{li05}, who performed global SPH + $N$-body simulations and
found that a critical ``combined star-gas'' Toomre parameter of 
$Q_{sg}=1.6$ defines the limit between gravitationally-stable and
unstable disks.  Of course, turbulence driven
by swing amplification could change the effective value of $Q_g$ if 
it contributes a stabilizing pressure in a manner similar to the
microphysical random motions, and if the turbulent amplitudes are
large. Since the threshold radii are by
definition where disks become stable, however, then 
$Q_g >\Qgcrit$ in those locations and there would not be a 
significant contribution from swing-driven turbulence.

We note that many physical processes not included in the
current models may potentially change $\Qgcrit$ for gravitational
runaway.  These include turbulence pervasive in the ISM, and 
thermal and other dynamical instabilities. 
For instance, \citet{kim03} found that nonlinear density fluctuations
driven by magnetorotational instability in gas-only, three-dimensional disks 
enhance $\Qgcrit$ by about 50\% relative to those in unmagnetized counterparts.
It is especially important to explore the effects of turbulent
motions and magnetic fields in multiphase systems, because the
thermal velocity dispersion of cold gas is so low.  The usual
assumption is that an effective $Q$ can be defined by appropriate
weighting of various contributing effective pressures and surface
density components (cf. \citealt{wad02}); an important direction for
future research is to test this idea rigorously.\footnote{Single-phase
models such as the present one are not useful for evaluating the idea or 
quantifying the value of $Q_{g, \rm eff}$, since naive inclusion of the 
radial turbulent velocities that are present ($< 0.5c_g$) at both small 
and large scales increases $Q_g$ by $<10\%$.} 
In future work, it will be interesting to test whether realistic 
turbulent, multi-phase models (in 3D or with thick-disk 2D gravity)
show nonlinear threshold behavior for a suitably defined $Q_{g,\rm eff}$,
and how a live stellar component quantitatively affects critical $Q$ values.

The fact that the stellar disk plays a significant role in
destabilizing the gas disk may help explain the very low star
formation rates in late-type, low surface brightness (LSB) galaxies.
 While LSB galaxies are
often comparable in total gas mass to normal galaxies (e.g.,
\citealt{mat05}), most of them have gas surface densities below the
threshold values $\Sigma_{g,\rm th}$ corresponding to $Q_g=1.4$, such
that low star formation rates and weak stellar disks are easily
explained (e.g., \citealt{van93,deb96,uso03}).  However, some LSB
galaxies with $Q_g<1.4$ still evidence little
star formation (\citealt{pic97,pic99,one00-1,one00-2}; see also
\citealt{elm02}).  In part, this is probably because $Q_s$ is so large
that the stellar disk does little to encourage instability in the gas
disk; we found that pure-gas disks are still quite stable when $Q_g=1.2$.
In addition, without the vertical gravity of a massive stellar disk,
the gas disk thickness will increase by a factor of 
1.5--2 (e.g. \citealt{kos02}) compared to that in a normal galaxy.
A larger value of $H_g$ dilutes gravity, which tends to further lower 
$\Qgcrit$.  Together, these effects will reduce $\Qgcrit$ below 1 for
LSB galaxies, making them more stable than previously thought.

Galactic differential rotation represents a bountiful store of kinetic
energy (e.g., \citealt{vonw51}), and we have shown that swing
amplification in two-component disks is able to transform this energy
into ISM turbulent motions with appreciable amplitudes.  Other
mechanisms that can channel sheared rotation into turbulence in the diffuse
ISM include the magnetorotational instability
\citep{sel99,kim03,pio04,pio05}, and
interactions with spiral arms (\citealt{mar98,gom04,kim06,kko06}).  
Our current simulations show that provided that the disk is marginal or
unstable, swing amplification is at least as efficient
as the magnetorotational instability under conditions of comparable
surface density.   Two-phase self-gravitating hydrodynamic models
\citep{wad02} and self-gravitating dissipative cloud-fluid models 
\citep{hub01} have similarly concluded that self-gravity 
coupled with galactic rotation can produce ISM turbulence that is
sustained over many galactic rotations.

Driving of turbulence at levels $\sigma_v \simgt 0.6 c_g$
requires a disk to be at least marginally unstable.  Because the outer,
low-surface density regions of disks are quite stable, this mechanism cannot
drive turbulence there.\footnote{The outer parts of disks, however, are also
where the scale height flares; together with low \ion{H}{1} surface
density this implies low gas volume densities, which is most favorable
for the magnetorotational instability to develop (Piontek \& Ostriker
2006, in preparation).} But in the inner parts of disks, self-gravitational
stirring in some regions (having slightly lower density) may go
hand-in-hand with formation of massive, star-forming clouds
in other regions (having slightly higher density).  Because the
fraction of gas converted to stars per cloud formation epoch is quite
small ($\simlt 10$\%), the rate of reduction in the mean gas surface density --
and hence rate of increase in $Q_g$ -- is very low.  As $Q_g$
increases over time, the timescale for cloud formation via 
gravitational instability also increases.  The overall evolution is 
toward an asymptotic state in which $Q_g$ approaches $\Qgcrit$.
Galaxies may naturally ``fine-tune'' their parameters in such a
way that self-gravity can help stabilize small scales -- by generating
turbulence -- even as it forces large scales into collapse.

\acknowledgments

W.-T.~K. gratefully acknowledges the assistance and hospitality provided 
by the Department of Astronomy at the University of Maryland while 
this paper was prepared. W.-T.~K. is supported by Korea Science and Engineering
Foundation (KOSEF) grant R01-2004-000-10490-0 at SNU.  This work is
partially supported by NASA grant NNG05GG43G to the University of Maryland.

%fig1 

%fig2

%fig3

%fig4

%fig5

%fig6

%fig7

%fig8

%fig9

%fig10

%fig11

%fig12

%fig13


\begin{thebibliography}{}
\bibitem[Balbus \& Cowie(1985)]{bal85}
   Balbus, S.\ A., \& Cowie, L.\ L. 1985, \apj, 297, 61
\bibitem[Barbanis \& Woltjer(1967)]{bar67}
   Barbanis, B., \& Woltjer, L.\ 1967, 150, 461
\bibitem[Binney \& Tremaine(1987)]{bin87} 
   Binney, J., \& Tremaine, S.\ 1987, 
   Galactic Dynamics (Princeton: Princeton Univ.\ Press)
\bibitem[Boldyrev et al.(2002)]{bol02}
   Boldyrev, S., Nordlund, \AA., \& Padoan, P.\ 2002, \apj, 573, 678
\bibitem[Bottema(1993)]{bot93}
   Bottema, R.\ 1993, \aap, 275, 16
\bibitem[Boulares \& Cox(1990)]{bou90}
   Boulares, A., \& Cox, D.\ P.\ 1990, \apj, 365, 544
\bibitem[Caldwell et al.(1991)]{cal91}
   Caldwell, N., Kennicutt, R., Phillips, A.\ C., \& Schommer, R.\ A.\
   1991, \apj, 370, 526
\bibitem[Carlberg \& Sellwood(1985)]{car85}
   Carlberg, R.\ G., \& Sellwood, J.\ A.\ \apj, 292, 79
\bibitem[Chen et al.(2001)]{che01}
   Chen, B., Stoughton, C., Smith, A.\ et al.\ 2001, \apj, 553, 184
\bibitem[de Blok et al.(1996)]{deb96}
   de Blok, W.\ J.\ G., McGaugh, S.\ S., \& van der Hulst, J.\ M.\
   1996, \mnras, 283, 18
\bibitem[de Grijs \& Peletier(1997)]{deg97} de Grijs, R., \& 
Peletier, R.~F.\ 1997, \aap, 320, L21
\bibitem[De Simone et al.(2004)]{des04}
   De Simone, R., Wu, X., \& Tremaine, S.\ 2004, \mnras, 350, 627
\bibitem[Dickey \& Lockman(1990)]{dic90} Dickey, J.\ M., \&
   Lockman, F.\ J.\ 1990, \araa, 28, 215
\bibitem[Dickey et al.(1990)]{dic_etal90} Dickey, J.\ M., Hanson, M.\ M.,
   \& Helou, G.\ 1990, \apj, 352, 522 
\bibitem[Elmegreen(1987)]{elm87} Elmegreen, B.\ G.\ 1987, \apj, 312, 626
\bibitem[Elmegreen(1995a)]{elm95a} Elmegreen, B.\ G.\ 1995a,
   in The 7th Guo Shoujing Summer School on Astrophysics:
   Molecular Clouds and Star Formation, eds.\ C.\ Yuan \&
   Hunhan You (Singapore:World Scientific), 149
\bibitem[Elmegreen(1995b)]{elm95b} Elmegreen, B.\ G.\ 1995b, \mnras, 275, 944
\bibitem[Elmegreen(2002)]{elm02}
   Elmegreen, B.\ G.\ 2002, \apj, 577, 206
\bibitem[Elmegreen \& Elmegreen(1983)]{elm83}
   Elmegreen, B.\ G., \& Elmegreen, D.\ M.\ 1983, \mnras, 203, 31
\bibitem[Elmegreen et al.(2003)]{elm03} Elmegreen, B.~G., 
Elmegreen, D.~M., \& Leitner, S.~N.\ 2003, \apj, 590, 271 
\bibitem[Elmegreen \& Scalo(2004)]{elm04}
    Elmegreen, B.\ G., \& Scalo, J.\ 2004, \araa, 42, 211
\bibitem[Fuchs(2001)]{fuc01}
   Fuchs, B.\ 2001, \mnras, 325, 1637
\bibitem[Fuchs(2005)]{fuc05}
   Fuchs, B., Dettbarn, C., \& Tsuchiya, T.\ 2005, \aap, 444, 1
\bibitem[Gammie(1996)]{gam96} Gammie, C.\ F.\ 1996, \apj, 462, 725
\bibitem[Gammie(2001)]{gam01} Gammie, C.\ F.\ 2001, \apj, 553, 174
\bibitem[Goldreich \& Lynden-Bell(1965a)]{gol65a} Goldreich, 
   P., \& Lynden-Bell, D.\ 1965a, \mnras, 130, 97
\bibitem[Goldreich \& Lynden-Bell(1965b)]{gol65b}
   Goldreich, P., \& Lynden-Bell, D.\ 1965b, \mnras, 130, 125
\bibitem[G\'omez \& Cox(2004)]{gom04}
   G\'omez, G.\ C., \& Cox, D.\ P.\ 2004, \apj, 615, 744
\bibitem[Griv et al.(1999)]{gri99}
   Griv, E., Rosenstein, B., Gedalin, M., \* Eichler, D.\
   1999, \aap, 347, 821
\bibitem[Hawley, Gammie, \& Balbus(1995)]{haw95}
   Hawley, J.\ F., Gammie, C.\ F., \& Balbus, S.\ A.\ 1995, \apj, 440, 742
\bibitem[Heiles(2001)]{hei01}
   Heiles, C.\ 2001, in ASP Conf.\ Ser.\ 231, Tetons 4:
      Galactic Structure, Stars, and the Interstellar Medium, eds.\
      C.\ E.\ Woodward,
      M.\ D.\ Bicay \& J.\ M.\ Shull (ASP: San Francisco), 294
\bibitem[Hockney \& Eastwood(1988)]{hoc88}
   Hockney, R.\ W., \& Eastwood, J.\ W.\ 1988, Computer Simulation
   Using Particles (Philadelphia: Adam Hilger)
\bibitem[Holmberg \& Flynn(2000)]{hol00}
   Holmberg, J., \& Flynn, C.\ 2000, \mnras, 313, 209
\bibitem[Huber \& Pfenniger(2001)]{hub01}
   Huber, D., \& Pfenniger, D. 2001, \aap, 374, 465
\bibitem[Hunter(1964)]{hun64} Hunter, C.\ 1964, \apj, 139, 570
\bibitem[Jog(1992)]{jog92} Jog, C.\ J.\ 1992, \apj, 390, 378
\bibitem[Jog(1996)]{jog96} Jog, C.\ J.\ 1996, \mnras, 278, 209
\bibitem[Jog \& Solomon(1984a)]{jog84a}
   Jog, C.\ J., \& Solomon, P.\ M.\ 1984a, \apj, 276, 114
\bibitem[Jog \& Solomon(1984b)]{jog84b}
   Jog, C.\ J., \& Solomon, P.\ M.\ 1984b, \apj, 276, 127
\bibitem[Julian \& Toomre(1966)]{jul66}
   Julian, W.\ H., \& Toomre, A.\ 1966, \apj, 146, 810 
\bibitem[Karaali et al.(2004)]{kar04}
   Karaali, S., Bilir, S., \& Hamzao\u{g}lu, E.\ 2004, \mnras, 355, 307
\bibitem[Kennicutt(1989)]{ken89} Kennicutt, R.\ C.\ 1989, \apj, 344, 685
\bibitem[Kennicutt(1997)]{ken97} Kennicutt, R.\ C.\ 1997,
   in The Interstellar Medium in Galaxies,
   ed.\ J.\ M.\ van der Hulst (Dordrecht:Kluwer), 171
\bibitem[Kim \& Ryu(2005)]{jkim05}
   Kim, J., \& Ryu, D.\ 2005, \apj, 630, L45
\bibitem[Kim, Kim, \& Ostriker(2006)]{kko06} Kim, C.-G., Kim, W.-T.,
  and Ostriker, E.\ C. 2006, \apj, in press
\bibitem[Kim \& Ostriker(2001)]{kim01}
   Kim, W.-T., \& Ostriker, E.\ C.\ 2001, \apj, 559, 70 (Paper I)
\bibitem[Kim, Ostriker, \& Stone(2002)]{kos02}
   Kim, W.-T., Ostriker, E.\ C., \& Stone, J.\ M.\ 2002, \apj, 581, 1080
\bibitem[Kim, Ostriker, \& Stone(2003)]{kim03}
   Kim, W.-T., Ostriker, E.\ C., \& Stone, J.\ M.\ 2003, \apj, 599, 1157
\bibitem[Kim \& Ostriker(2006)]{kim06}
   Kim, W.-T., \& Ostriker, E.\ C.\ 2006, \apj, 646, 213
\bibitem[Knapen et al(1993)]{kna93}
   Knapen, J.\ H., Cepa, J., Beckman, J.\ E., Soledad del Rio, M.,
   \& Pedlar, A.\  1993, \apj, 416, 563
\bibitem[Kuijken \& Gilmore(1989)]{kui89}
   Kuijken, K., \& Gilmore, G.\ 1989, \mnras, 239, 605
\bibitem[Li, Mac Low, \& Klessen(2005)]{li05}
   Li, Y., Mac Low, M.-M., \& Klessen, R.\ S.\ 2005, \apj, 620, L19
\bibitem[Li et al.(2006)]{li06} 
   Li, Y., Mac Low, M.-M., \& Klessen, R.~S.\ 2006, \apj, 639, 879 
\bibitem[Lin \& Shu(1966)]{lin66}
   Lin, C.\ C., \& Shu, F.\ H.\ 1966, Proc.\ Nat.\ Acad.\ Sci., 55, 229
\bibitem[Lin, Yuan, \& Shu(1969)]{lin69}
   Lin, C.\ C., Yuan, C., \& Shu, F.\ H.\ 1969, \apj, 155, 721
\bibitem[Lubow, Balbus, \& Cowie(1986)]{lub86}
   Lubow, S.\ H., Balbus, S.\ A., \& Cowie, L.\ L.\ 1986, \apj, 309, 496
\bibitem[Mac Low \& Klessen(2004)]{mac04}
   Mac Low, M.-M., \& Klessen, R.\ S.\ 2004, Rev.\ Mod.\ Phys.\ 2004, 76, 125
\bibitem[Martin \& Kennicutt(2001)]{mar01}
   Martin, C.\ L., \& Kennicutt, R.\ C.\ 2001, \apj, 555, 301
\bibitem[Martos \& Cox(1998)]{mar98}
   Martos, M.\ A., \& Cox, D.\ P.\ 1998, \apj, 509, 703
\bibitem[Matthews et al.(2005)]{mat05}
   Matthews, L.\ D., Gao, Y., Uson, J.\ M., \& Combes, F.\ 2005, 
   \aj, 129, 1849
\bibitem[Minchev \& Quillen(2006)]{min06}
   Minchev, I., \& Quillen, A.\ C.\ 2005, \mnras, 368, 623 
\bibitem[Monaghan(1989)]{mon89}
   Monaghan, J.\ J.\ 1989, J.\ Comp.\ Phys., 82, 1
\bibitem[O'Neil, Bothun, \& Schombert(2000)]{one00-1}
   O'Neil, K., Bothun, G.\ D., Schombert, J.\ 2000, \aj, 119, 136
\bibitem[O'Neil, Verheijen, \& McGaugh(2000)]{one00-2}
   O'Neil, K., Verheijen, M.\ A.\ W., McGaugh, S.\ S.\ 2000, \aj, 119, 2154
\bibitem[Pickering et al.(1997)]{pic97}
   Pickering, T.\ E., Impey, C.\ D., van Gorkom, J.\ H., \& 
   Bothun, G.\ D.\ 1997, \aj, 114, 1858
\bibitem[Pickering et al.(1999)]{pic99}
   Pickering, T.\ E., van Gorkom, J.\ H., Impey, C.\ D., \& 
   Quillen, A.\ C.\ 1999, \aj, 118, 765
\bibitem[Piontek \& Ostriker(2004)]{pio04}
   Piontek, R.\ A., \& Ostriker, E.\ C.\ 2004, \apj, 601, 905
\bibitem[Piontek \& Ostriker(2005)]{pio05}
   Piontek, R.\ A., \& Ostriker, E.\ C.\ 2005, \apj, 629, 849
\bibitem[Quirk(1972)]{qui72} Quirk, W.\ J.\ 1972, \apj, 176, L9
\bibitem[Rafikov(2001)]{raf01}
   Rafikov, R.\ R.\ 2001, \mnras, 323, 445
\bibitem[Romeo(1992)]{rom92}
   Romeo, A.\ B.\ 1992, \mnras, 256, 307
\bibitem[Rybicki(1971)]{ryb71}
   Rybicki, G.\ B.\ 1971, \apss, 14, 15
\bibitem[Sellwood \& Carlberg(1984)]{sel84}
   Sellwood, J.\ A., \& Carlberg, R.\ G.\ 1984, \apj, 282, 61
\bibitem[Sellwood and Balbus(1999)]{sel99}
   Sellwood, J.\ A., \& Balbus, S.\ A.\ 1999, \apj, 511, 660
\bibitem[Stone \& Norman(1992)]{sto92}
   Stone, J.\ M., \& Norman, M.\ L.\ 1992, \apjs, 80, 753
\bibitem[Toomre(1964)]{too64} Toomre, A.\ 1964, \apj, 139, 1217
\bibitem[Toomre(1981)]{too81} Toomre, A.\ 1981, in
   Structure and Evolution of Normal Galaxies, eds.\ S.\ M.\ Fall
   \& D.\ Lynden-Bell (Cambridge:Cambridge Univ.\ Press), 111
\bibitem[Uson \& Matthews(2003)]{uso03}
   Uson, J.\ M., \& Matthews, L.\ D.\ 2003, \aj, 125, 2455
\bibitem[van der Hulst et al.(1993)]{van93} van der Hulst, J.\ M.,
   Skillman, E.\ D., Smith, T.\ R., Bothun, G.\ D., McGaugh, S.\ S.,
   \& de Blok, W.\ J.\ G.\ 1993, \aj, 106, 548
\bibitem[van der Kruit \& Searle(1981)]{van81} van der Kruit, 
P.~C., \& Searle, L.\ 1981, \aap, 95, 105
\bibitem[van der Kruit \& Searle(1982)]{van82} van der Kruit, 
P.~C., \& Searle, L.\ 1982, \aap, 110, 61
\bibitem[van Zee \& Bryant(1999)]{vanz99}
   van Zee, L., \& Bryant, J.\ 1999, \aj, 118, 2172
\bibitem[Vestuto et al.(2003)]{ves03} Vestuto, J.~G., 
Ostriker, E.~C., \& Stone, J.~M.\ 2003, \apj, 590, 858 
\bibitem[von Weizs\"acker(1951)]{vonw51}
   von Weizs\"acker, C.\ F.\ 1951, \apj, 114, 165
\bibitem[Wada, Meurer, \& Norman(2002)]{wad02}
   Wada, K., Meurer, G., \& Norman, C.\ A.\ 2002, \apj, 577, 197
\bibitem[White(1988)]{whi88}
   White, R.\ L.\ 1988, \apj, 330, 26
\bibitem[Wisdom \& Tremaine(1988)]{wis88}
   Wisdom, J., \& Tremaine, S.\ 1988, \aj, 95, 925
\bibitem[Wong \& Blitz(2002)]{won02}
   Wong, T., \& Blitz, L.\ 2002, \apj, 569, 157



\end{thebibliography}
\end{document}